\documentclass[a4paper, notoc]{JHEP3}
\usepackage[dvips]{graphicx}
\usepackage{amsfonts}
\usepackage{amssymb}
\usepackage{epsfig}
\usepackage{cite}

\newcommand{\be}{\begin{equation}}
\newcommand{\ee}{\end{equation}}
\newcommand{\bea}{\begin{eqnarray}}
\newcommand{\eea}{\end{eqnarray}}
\newcommand{\mbb}{\mathbb}
\newcommand{\ti}{\times}
\newcommand{\half}{\frac{1}{2}}
\newcommand{\mc}{\mathcal}

\newcommand{\lsim}{\lesssim}

\newcommand{\beqa}{\begin{eqnarray}}
\newcommand{\eeqa}{\end{eqnarray}}

\newcommand{\MET}{{\not\!\!p_T}}
\newcommand{\Dphi}{\Delta\phi}

 \newlength{\wth}
 \newcommand{\twographs}[2]{%
 \unitlength=1.1in
 \begin{picture}(5.8,2)
 \put(0,0){\epsfig{file=#1.eps, width=0.8\wth}}
 \put(2.5,0){\epsfig{file=#2.eps, width=0.8\wth}}
 \put(0,2.0){(a)}
 \put(2.6,2.0){(b)}
 \end{picture}
}

\newcommand{\threegraphs}[3]{%
 \unitlength=1.1in
 \begin{picture}(6.,5.2)
 \put(1.1,0.4){\epsfig{file=#3, width=0.75\wth}}
 \put(-0.2,2.3){\epsfig{file=#1, width=0.75\wth}}
 \put(2.5,2.3){\epsfig{file=#2, width=0.75\wth}}
 \put(1.1,2.1){(c)}
 \put(0,4.3){(a)}
 \put(2.2, 4.3){(b)}
\end{picture}}

 \title{
Sparticle
 Spectra and LHC Signatures 
   for Large Volume String Compactifications}
\author{J.~P. Conlon, C.~H. Kom, K. Suruliz,
  B.~C. Allanach, F. Quevedo\\ DAMTP,
  Centre for Mathematical
Sciences,
  Wilberforce Road, Cambridge, CB3 0WA, United Kingdom\\
 E-mail:
  \email{j.p.conlon@damtp.cam.ac.uk}, \email{c.kom@damtp.cam.ac.uk},
  \email{k.suruliz@damtp.cam.ac.uk},  \email{b.c.allanach@damtp.cam.ac.uk},
  \email{f.quevedo@damtp.cam.ac.uk}
}

\abstract{We study the supersymmetric particle spectra and LHC collider
observables for the
large-volume string models
with a fundamental scale of $10^{11} \hbox{~GeV}$
that arise in moduli-fixed string compactifications with branes and fluxes.
The presence of magnetic fluxes on the brane world volume, required for
chirality, perturb
the soft terms away from those previously computed in the dilute-flux
limit. We use the difference in
high-scale gauge couplings to estimate the magnitude of this perturbation
and study the potential effects of the magnetic fluxes by generating
many random spectra with the soft terms perturbed
around the dilute flux limit.
Even with a $40 \%$ variation in the high-scale soft terms the low-energy
spectra take a clear and predictive form.
The resulting spectra are broadly similar to those arising on the SPS1a slope,
but
more degenerate.
In their minimal version the models predict the ratios of gaugino masses to be
$M_1 : M_2 : M_3=(1.5 - 2) : 2 : 6$,
different to both mSUGRA and mirage
mediation. Among the scalars, the squarks tend to be lighter and the sleptons heavier than for comparable mSUGRA models.
We generate $10 \hbox{fb}^{-1}$ of sample LHC data for
the random spectra
in order to study the range of collider phenomenology that can occur.
We perform a detailed  mass reconstruction on one example large-volume string
model spectrum.
$100 \hbox{fb}^{-1}$ of integrated luminosity is sufficient to discriminate the
model from mSUGRA and aspects of the sparticle spectrum can be accurately
reconstructed.
}

\preprint{DAMTP-2007-33}

\begin{document}

\tableofcontents

\section{Introduction}

The imminent advent of the
Large Hadron Collider (LHC) is an excellent motivation to develop
techniques to relate high energy string compactifications to
observable low-energy collider physics. The LHC will be an
unprecedented probe of the terascale and of the physics that
stabilises the electroweak hierarchy. If supersymmetry is discovered
at the LHC, it will be necessary to connect the collider observables
and the spectrum of superparticles to a more fundamental theory such
as string theory.

Supersymmetric phenomenology is the study not of supersymmetry but
of supersymmetry breaking: the undetermined parameters of the
minimal supersymmetric standard model (MSSM)
are those associated with the soft supersymmetry breaking terms,
such as gaugino masses, scalar masses or trilinear A-terms. There
are over one hundred such parameters in a general phenomenological
parameterisation of the
MSSM. The many different
possibilities for supersymmetric phenomenology are determined by the
many different possibilities for the soft terms. However, high scale constructions such as
string compactifications contain far fewer independent parameters and so
could
be expected to lead to distinctive patterns of soft terms.

Broken supersymmetry is a property of the vacuum of the theory. The study of supersymmetry breaking
vacua in string theory therefore requires control over both the moduli potential and the
quantum corrections that enter into it. The quantum corrections play a crucial role for the
large-volume models \cite{hepth0502058, hepth0505076}
that we study here.
The first
constructions of IIB string models with all moduli stabilised
\cite{hepth0301240} involved unbroken supersymmetry, which had to be broken by
hand
by the addition of
$\overline{{\rm D3}}$-branes, with a gravitino mass that could only be lowered below the Planck scale by fine-tuning.
It was subsequently realised that with the inclusion of quantum corrections \cite{hepth0204254}
the moduli potential can admit
non-supersymmetric minima \cite{hepth0408054} and even naturally generate a hierarchically small
supersymmetry-breaking scale \cite{hepth0502058, hepth0505076}. This
is the large volume scenario that we will explore here.\footnote{The possibility of
generating low-scale supersymmetry breaking from the moduli-stabilising potential has also been
studied in corners of string theory different from that of IIB with
fluxes \cite{bobby}.}

Nonetheless, it is still a long way from the mere existence of
low-scale supersymmetry breaking to actual phenomenology, and
turning any given string compactification into LHC
collider observables requires many steps. We will
assume gravity-mediated supersymmetry breaking as the most directly motivated scenario
from string compactifications; the modifications
 for other proposals are straightforward.

\begin{enumerate}

\item The first requirement is that the compactification moduli be
stabilised within a controlled approximation. This is necessary to
ensure that the compactification is a good vacuum solution of string
theory.

\item The second requirement is that supersymmetry is broken in the
vacuum, so that soft terms can be generated. We also require the
supersymmetry breaking to be hierarchically small, in order that the
soft terms appear at the TeV scale rather than near the Planck scale.

\item The third requirement is a visible sector containing the
Standard Model gauge group, with an understanding of the
non-renormalisable couplings between the visible sector and the
hidden sector moduli that break supersymmetry.

\item The fourth task is to combine these couplings with the hidden
  sector supersymmetry
breaking to compute the visible sector soft supersymmetry breaking
terms at the high-energy compactification scale. Here we also
would like an understanding of why the soft terms generated do not lead
to large CP violation or flavour-changing neutral currents.

\item The fifth task is to run these soft terms down to the
TeV-scale, in order to compute the physical sparticle spectrum.

\item The sixth task is to put such a spectrum through event
generators such as \cite{pythia, herwig} and detector simulators
such as \cite{PGS} in order to generate collider observables.

\item Finally, we also want an estimate of the uncertainties
arising from the above six steps.

\end{enumerate}

The aim of this paper is to carry out this program almost in its entirety
for a specific and well-motivated class of string compactifications,
the large volume models
developed in ref.~\cite{hepth0502058}.\footnote{The omission
  is in the lack of
an explicit magnetised brane configuration realising the
MSSM gauge group;
this we simply assume can be
achieved.}  These models arise within IIB flux
compactifications with moduli stabilisation and are characterised by broken supersymmetry with an
exponentially large compactification volume. This allows the natural
generation of hierarchies
between mass scales,
an extremely desirable feature. The large volume $\mc{V}$
lowers both the string scale $m_s$ and
the gravitino mass $m_{3/2}$ with respect to the Planck scale $M_P$,
\begin{equation}
\label{fe}
m_{s} \sim \frac{M_P}{\sqrt{\mc V}} \, , \qquad m_{3/2} \sim
\frac{M_P}{\mc V}. \label {volrelns}
\end{equation}
$\mc V$ refers to a dimensionless quantity: the volume measured in powers of
the string-length $l_s$. The dimensionful volume of the compactification manifold is ${\mc V} l_s^6$.
From (\ref{fe}), an intermediate string scale $m_s \sim 10^{11} \hbox{~GeV}$ gives rise to
TeV-scale supersymmetry breaking. The phenomenological implications of
these models have been studied in \cite{hepth0502058, hepth0505076, hepph0512081,
   hepth0605141, hepth0610129} where requirements 1-5 above have been
addressed. In this article we will first review
the
most relevant results from those references and complete requirements
6 and 7 above in order to make direct contact with potential
observables at the LHC.

 The organisation of this paper is as follows. In section \ref{largevol} we review
the large-volume models and the moduli stabilisation that generates the exponentially large volume.
We also describe the computation of soft terms and explain why leading order flavour universality is assured,
summarising the results of refs.~\cite{hepth0502058, hepth0505076,
  hepth0609180,
hepth0610129}.
We also describe how magnetic fluxes perturb the soft terms away from those computed in the dilute-flux limit.
We estimate the magnitude of the flux perturbation from the non-universality of the high-scale gauge couplings.
Such corrections will generate an uncertainty in the high-scale soft terms that will translate into
an uncertainty in the low-energy spectra and observables.
In section \ref{spectra} we
examine the low-energy spectra arising from the soft terms
considered in section \ref{softbreak}, by generating random soft terms perturbed about the dilute-flux
limit. We describe the generic
properties of the resulting low-energy spectra and compare with those arising in mSUGRA or mirage mediation
models. In section \ref{observables} we study collider observables for these spectra.
We use counting observables to scan the properties of the randomly generated spectra, and
show that for a sample spectrum sparticle masses can be reconstructed. In section \ref{conclusions} we present our conclusions.

We have made an effort to make this article self-contained. A
phenomenologically minded reader may wish to skip the formal details of
section 2 and start in section 2.3.

\section{Large-Volume Models}
\label{largevol}

Large-volume models represent a class of string compactifications, with
all moduli stabilised, in which quantum corrections to the
scalar potential naturally lead to a exponentially large volume.
They were first found in references \cite{hepth0502058, hepth0505076}.
They are robust against additional quantum corrections, such as those of \cite{hepth0507131, hepth0508043}.
A recent detailed study of this robustness is
\cite{berg}. They have been applied to obtain string theory inflation
\cite{inflation}, natural QCD axions \cite{hepth0602233}, the scale for
neutrino masses \cite{hepph0611144}
and to
low
 energy  supersymmetric phenomenology  \cite{hepph0512081,hepth0605141,hepth0609180, hepth0610129} in which they provide a natural
 hierarchy for supersymmetry breaking with the large volume leading to
 an intermediate string scale. A comprehensive review is \cite{review}.
 We start by reviewing their construction and properties.


\subsection{Construction}

These models arise as a rather generic limit of flux
compactifications of IIB string theory in the presence of D3 and D7
branes. In $N=1$ supersymmetric IIB compactifications the K\"ahler
potential and superpotential for the moduli
$\Phi=S, U_a, T_i$ take the
standard form \cite{hepth0301240, hepth9906070, hepth0204254,
hepth0403067},
\bea
\label{KahlerPot}
\hat{K}(\Phi, \bar{\Phi}) & = & - 2 \ln \left(
\mc{V} + \frac{\hat{\xi}}{2 g_s^{3/2}} \right) - \ln \left(
i \int \Omega \wedge \bar{\Omega} \right) - \ln (S + \bar{S}), \\
\hat{W}(\Phi) & = & \int G_3 \wedge \Omega + \sum_i A_i e^{-a_i
T_i},
\eea
where the dependence on the complex structure moduli $U$ is encoded in
the Calabi-Yau $(3,0)$ form $\Omega$. $G_3$ corresponds to the
three-form fluxes and is linear in the dilaton $S$.
We have included the leading $\alpha'$ correction to the K\"ahler
potential, which depends on $\hat{\xi} = - \zeta(3) \chi(M)/(2 \pi)^3$
with $\chi(M)$ the Euler number of the Calabi-Yau manifold $M$.
Large-volume models require $M$ to have at
least two K\"ahler moduli $T_i$, one of which is a blow-up mode, as well as a negative
Euler number,
i.e.\ $\chi(M) < 0$.
The simplest model is that of $\mbb{P}^4_{[1,1,1,6,9]}$, which we
use as our working example. For this the volume can be written as
\cite{hepth0502058, hepth0404257} \be \mc{V} = \frac{1}{9\sqrt{2}} \left(
\tau_b^{3/2} - \tau_s^{3/2} \right). \ee $\tau_b = \hbox{Re}(T_b)$ and $\tau_s = \hbox{Re}(T_s)$
denote big and small cycles. The geometry is analogous to that of a Swiss cheese: the cycle $T_b$ controls the volume
(`the size of the cheese') and $T_s$ controls a blow-up cycle (`the size of the hole').

The $\mc{N}=1$ moduli scalar potential is \be
\label{n=1pot} V = e^{\hat{K}} \left( \hat{K}^{i \bar{j}} D_i W D_{\bar{j}}
\bar{W} - 3 \vert W \vert^2 \right), \ee
where $D_iW= \partial_i W +
(\partial_i \hat{K})W$.
Dropping terms sub-leading in $\mc{V}$, this potential becomes
\be
\label{pot}
V = \sum_{\Phi = S,U} \frac{{\hat{K}}^{\Phi \bar{\Phi}} D_\Phi W \bar{D}_{\bar{\Phi}} \bar{W}}{\mc{V}^2}
+ \frac{\lambda (a_s A_s)^2 \sqrt{\tau_s} e^{-2 a_s \tau_s}}{\mc{V}} - \frac{\mu W_0 a_s A_s \tau_s e^{-a_s \tau_s}}{\mc{V}^2}
+ \frac{\nu \xi \vert W_0 \vert^2}{g_s^{3/2} \mc{V}^3}
\ee
in the limit $\mc{V} \gg 1$, where $\lambda, \mu, \nu$ are numerical constants.
The first terms of (\ref{pot})
 stabilise the dilaton and complex structure moduli at $D_S W = D_U W = 0$.
The remaining terms stabilise the K\"ahler moduli. The non-perturbative terms in $\tau_s$ balance against the
perturbative corrections in the volume, and it can be shown that at the minimum of the scalar
potential \cite{hepth0502058}
$$
\mc{V} \sim W_0 e^{\frac{c}{g_s}}, \qquad \tau_s \sim \ln \mc{V},
$$
where $W_0$ is the value of the flux superpotential at the minimum of
$S$ and $U$ fields, and $c \sim \xi^{2/3}$ is a numerical constant. The resulting volume is exponentially large with a small blow-up cycle.
This geometry is shown in figure \ref{LargeVolumePic}.
\FIGURE{\makebox[15cm]{\epsfxsize=15cm \epsfysize=12cm
\epsfbox{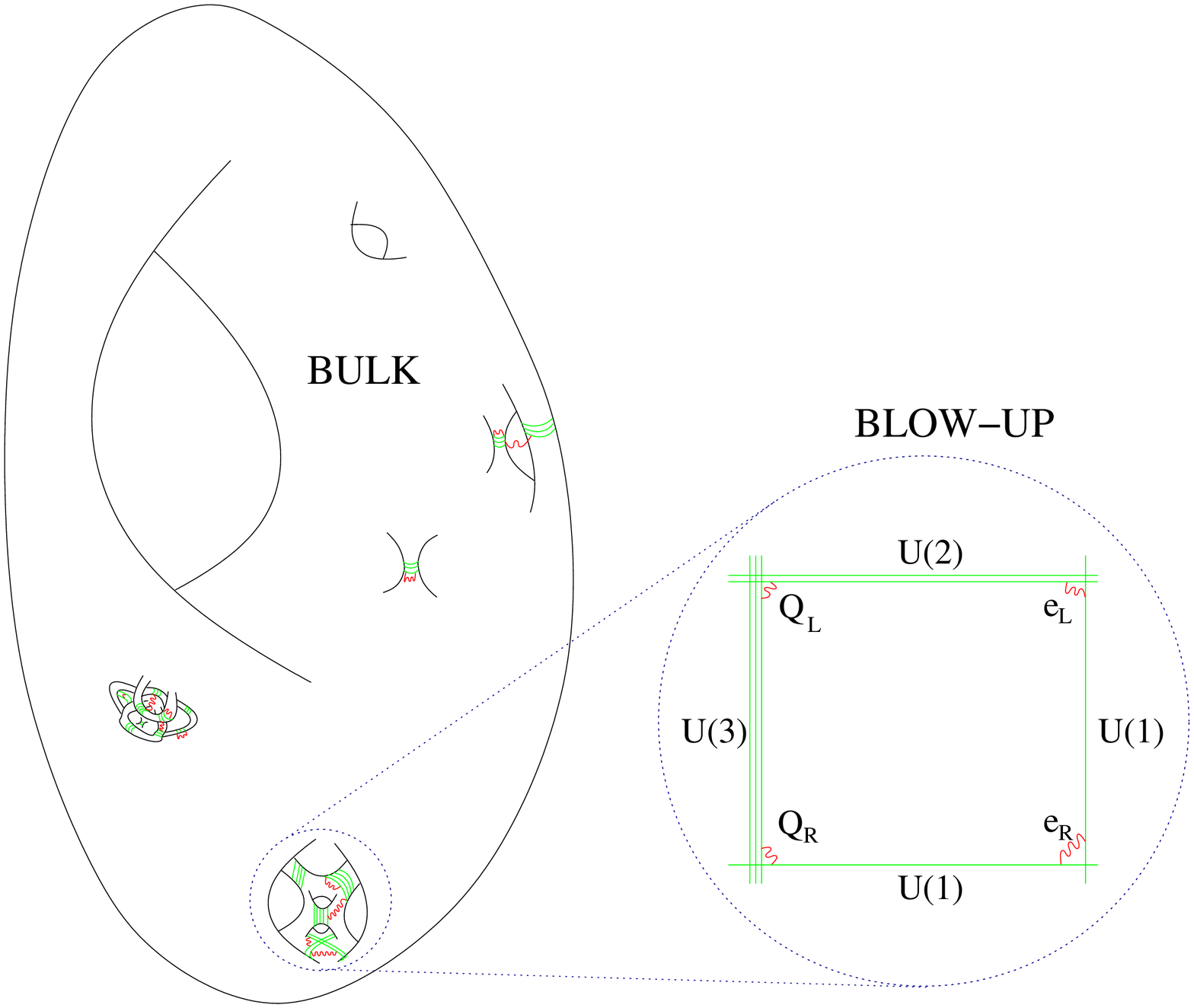}
}\caption{The physical picture:
Standard Model matter is supported on a small blow-up cycle located
within the bulk of a very large Calabi-Yau. The volume of the
Calabi-Yau sets the gravitino mass and is responsible for the
weak/Planck hierarchy.}\label{LargeVolumePic}}
The minimum avoids two problems of the KKLT scenario. First, as its consistency does not require a fine-tuning of $W_0$,
its existence is more generic.
Secondly, the stabilisation of the $T$ moduli can never destabilise the $U$ moduli, as
near the minimum the relevant terms in (\ref{pot})
come with different powers of the volume. A phenomenological advantage is that this minimum also breaks supersymmetry
and generates an exponentially large volume.\footnote{The minimum of
  the potential
  (\ref{pot}) is at negative vacuum energy $|V_0|\sim {m_{3/2}^3 M_P}$ with supersymmetry broken. Just as in KKLT a lifting
  term is desirable to obtain de Sitter or Minkowski space.
There are several possible sources for this lifting term
  \cite{bkq} (see \cite{daniel} for a recent detailed analysis for
  KKLT and large
  volume models). Since the anti de Sitter minimum is already
  non supersymmetric,
in contrast to KKLT models, the contribution of this
  lifting term to the soft breaking terms is suppressed and does not
  play an important role in the rest of this paper.}

This large volume allows the generation of hierarchies by lowering both the
string  scale and gravitino mass as in (\ref{volrelns}).
A volume $\mc{V} \sim 10^{15} $ is required to explain the weak/Planck
hierarchy and give TeV-scale supersymmetry breaking. We assume such a volume throughout this paper.
Further attractive features are that the volume $\mc{V} \sim 10^{15}$ also
gives an axion decay constant $f_a \sim 10^{11} \hbox{~GeV}$ in the allowed window \cite{hepth0602233} and the
required neutrino suppression scale of $10^{14} \hbox{~GeV}$ \cite{hepph0611144}.

A fully realistic model must have a Standard Model sector, which requires an appropriate configuration of
O-planes and magnetised D7-branes. Chirality arises from the topological intersection numbers of
differently magnetised branes.\footnote{Magnetic flux on D7 branes is equivalent to dissolved D5 branes. The chirality
arises from the point like intersection of the dissolved D5 branes.}
 To avoid having gauge groups that are too weakly coupled,
we must wrap the Standard Model D7 branes on the small cycle corresponding to $\tau_s$.
As the branes should wrap a blow-up cycle, the brane configuration will represent a
local construction of the Standard Model. We assume such a configuration can be found, but do not attempt to realise the
brane configuration explicitly.
The techniques involved
in explicitly constructing such a configuration will be similar to those used in
models of branes at singularities.
In this context there has been recent effort at
constructing MSSM-like gauge groups - see \cite{hepth0005067, hepth0610007, hepth0703047}
for progress.

As $m_s \gg m_{3/2}$, the infrared physics of these models is that of
the MSSM. The computation of the soft terms is then central to the
study of the low energy phenomenology.

\subsection{Gauge Couplings}

The gauge kinetic functions $f_a(\Phi)$ are principally determined by the cycles wrapped by the
D7 branes. If $T_i$ is the K\"ahler modulus
corresponding to a particular 4-cycle, reduction of the DBI action
for an unmagnetised brane wrapped on that cycle gives\footnote{This holds with the phenomenology
conventions $\hbox{Tr}(T^a T^b) = \half \delta^{ab}$; with conventions $\hbox{Tr}(T^a T^b) = \delta^{ab}$
the gauge kinetic function is $f = \frac{T}{2 \pi}$.}
\be
\label{GaugeCoupling} f_i = \frac{T_i}{4 \pi}.
\ee
To generate chirality we require the branes to be magnetised.
The magnetic fluxes alter
(\ref{GaugeCoupling}) to \be \label{FluxGaugeCouplings} f_i =
\frac{T_i}{4 \pi} + h_i(F) \, S, \ee where $h_i$ is a topological function of the fluxes
present on the brane. This can be understood microscopically. The
gauge coupling $1/g^2=\hbox{Re}  f$ 
 is given by an integral over the cycle $\Sigma$
wrapped by the brane, \be \frac{1}{g^2} = \int_{\Sigma} d^4 y \,
e^{-\phi} \sqrt{g + (2 \pi
  \alpha')\mc{F}}.
\ee
Thus in the presence of flux, the gauge coupling depends on the magnetic flux as well as
the cycle volume.\footnote{In computing explicit expressions for the factors $h_i(F)$,
it is easiest to use the Chern-Simons action to get the correction to $\hbox{Im}(f)$, and then use holomorphy for the correction
to $\hbox{Re}(T)$.}
The factors $h_i(F)$ have been explicitly computed for toroidal orientifolds
\cite{hepth0404134, hepth0406092}.

As a minimal scenario we assume
that
the Standard Model branes live on a single blow-up cycle and so
the gauge kinetic functions depend upon
only one K\"ahler modulus $T_s$.
As the cycle size $T$ is increased, the magnetic fluxes are diluted and
their contribution to the gauge couplings goes away. It is useful to consider
this dilute flux limit in addition to the physical case, and to see the latter as a
perturbation on the former. The Standard Model gauge kinetic functions
will be
\begin{enumerate}
\item Dilute flux limit
\bea \label{dilutef}
f_{SU(3)} & = & \frac{T_s}{4 \pi}, \nonumber \\
f_{SU(2)} & = & \frac{T_s}{4 \pi}, \nonumber \\
f_{U(1)_Y} & = & k_Y \frac{T_s}{4 \pi}.
\eea
\item Physical case
\bea
\label{f3}
f_{SU(3)} & = & \frac{T_s}{4 \pi} + h_{SU(3)}(F) S, \\
\label{f2}
f_{SU(2)} & = & \frac{T_s}{4 \pi} + h_{SU(2)}(F) S, \\
\label{f1}
f_{U(1)_Y} & = & k_{Y} \left( \frac{T_s}{4 \pi} + h_{U(1)}(F) S \right).
\eea
\end{enumerate}

The factor $k_Y$ accounts for the uncertainty in normalising the $U(1)$ gauge field: the $U(1) = U(n)/SU(n)$
that appears in intersecting brane models in general has a different normalisation to that of $U(1)_Y$.
The `canonical' value $k_Y = 5/3$ only holds for $SU(5)$ grand unified
models; in general, $k_Y$
is model dependent.
Typically in D-brane models
hypercharge is an anomaly free linear combination of different
$U(1)$'s coming from different $U(n)$ factors, the particular linear
combination determining the value of the normalisation factor $k_Y$.
The functions $h(F)$ depend on the microscopic configuration of branes and
fluxes, and we also regard these as unknown.

\subsection{Soft Breaking Terms}
\label{softbreak}

A four dimensional $N=1$ supergravity Lagrangian is specified at two
space-time derivatives by the K\"ahler potential $K$, superpotential $W$ and
gauge kinetic function $f_a$. The computation of soft terms starts
by expanding these as a power series in the matter fields, \bea
\label{PowerSeriesExpansion} W & = & \hat{W}(\Phi) + \mu(\Phi) H_1
H_2 + \frac{1}{6} Y_{\alpha
  \beta \gamma}(\Phi) C^{\alpha} C^{\beta} C^{\gamma} + \ldots, \\
\label{MatterK} K & = & \hat{K}(\Phi, \bar{\Phi}) +
\tilde{K}_{\alpha \bar{\beta}} (\Phi, \bar{\Phi}) C^{\alpha}
C^{\bar{\beta}} + \left[ Z(\Phi,
  \bar{\Phi}) H_1 H_2 + h.c. \right] + \ldots, \\
  \label{gkff}
f_a & = & f_a(\Phi). \eea Here $\Phi$ denotes a generic modulus field and
$C^\alpha$ a generic matter field. In the MSSM
the $\mu$ and $Z$ terms apply only to the Higgs fields and so for
these we have written $H_1$ and $H_2$ explicitly.
Gravity-mediated supersymmetry breaking is then quantified through the
moduli F-terms, given by
\be
F^m = e^{\hat{K}/2} \hat{K}^{m \bar{n}}
D_{\bar{n}} \bar{\hat{W}}.
\ee
The relationships of the expansions (\ref{PowerSeriesExpansion}) to (\ref{gkff}) to the
soft supersymmetry breaking terms are given in full detail in ref.~\cite{hepph9707209}.

\subsubsection{Gaugino Masses}

The canonically normalised gaugino masses are given by \be
\label{CNGM} M_a = \half \frac{F^m \partial_m f_a}{ \hbox{Re} f_a}.
\ee
The gaugino masses that follow from (\ref{FluxGaugeCouplings})
are
\begin{enumerate}
\item Dilute flux limit
\be
\label{DiluteFluxGauginoMasses}
M_1 = M_2 = M_3 = \frac{F^s}{2 \tau_s} \equiv M.
\ee
\item Physical case
\bea
M_1 & = & \frac{F^s}{2 (\tau_s + 4 \pi h_1(F) \hbox{Re}(S))}, \nonumber \\
M_2 & = & \frac{F^s}{2 (\tau_s + 4 \pi h_2(F) \hbox{Re}(S))}, \nonumber \\
M_3 & = & \frac{F^s}{2 (\tau_s + 4 \pi h_3(F) \hbox{Re}(S))},
\label{PhysicalGauginoMasses}
\eea
\end{enumerate}
 where we write $M_i$ for $M_{SU(i)}$, and similarly for $h_a$ and $f_a$.

In the limit of large cycle volume, the flux becomes
dilute and the gaugino masses become
universal at the compactification scale where the soft parameters are
computed.
In the physical case, the gaugino masses are non-universal due to the
flux contribution. However for non-abelian gauge groups - i.e. for
the wino and gluino - it follows from (\ref{f3}) and (\ref{f2}) that the
fractional non-universality of gaugino masses is identical to that of the
gauge couplings:
\be
\label{m3m2ratio}
\frac{M_3}{M_2} \Bigg\vert_{m_s} = \frac{g_3^2}{g_2^2} \Bigg\vert_{m_s}.
\ee
Due to the factor $k_Y$ in the $U(1)$ gauge couplings, this relation
does not hold for $M_1$. Here we have
\be
\frac{M_3}{M_1}\Bigg\vert_{m_s} = \frac{g_3^2}{k_Y g_1^2}\Bigg\vert_{m_s},
\ee
where $k_Y$ is the unknown normalisation factor.

\subsubsection{Scalar Soft Terms}
For the case of diagonal matter field metrics,
$\tilde{K}_{\alpha\bar{\beta}} = \tilde{K}_\alpha
\delta_{\alpha\bar{\beta}}$ (no summation over $\alpha$), the scalar masses, A-terms and
B-term are given by \cite{hepph9707209}
\bea \label{SoftMassDiagFormula} m_\alpha^2  &
=  & (m_{3/2}^2 + V_0) - F^{\bar{m}}F^n \partial_{\bar{m}}
\partial_n \log \tilde{K}_\alpha. \\
\label{ATermFormula} A_{\alpha \beta \gamma} & = & F^m \left[
\hat{K}_m + \partial_m \log
  Y_{\alpha \beta \gamma} - \partial_m \log (\tilde{K}_\alpha \tilde{K}_\beta \tilde{K}_\gamma) \right]. \\
\label{BTermFormula} B \hat{\mu} & = & (\tilde{K}_{H_1}
\tilde{K}_{H_2})^{-\half} \Bigg\{  e^{\hat{K}/2} \mu \left( F^m
\left[\hat{K}_m + \partial_m \log \mu - \partial_m \log
  (\tilde{K}_{H1} \tilde{K}_{H2}) \right] - m_{3/2} \right) \nonumber \\
& &  + \left( 2 m_{3/2}^2 + V_0 \right) Z - m_{3/2}
\bar{F}^{\bar{m}} \partial_{\bar{m}} Z + m_{3/2} F^m  \left[
\partial_m Z -
Z \partial_m  \log (\tilde{K}_{H_1} \tilde{K}_{H_2})\right] - \nonumber \\
& & \bar{F}^{\bar{m}} F^n \left[ \partial_{m} \partial_n Z -
(\partial_{\bar{m}} Z) \partial_n \log (\tilde{K}_{H_1}
\tilde{K}_{H_2})\right] \Bigg\}.
\eea
As mentioned above, the tree-level  vacuum energy $V_0$ (in Planck units) is much smaller
than $m_{3/2}^2$ ($V_0/M_P^2\lsim m_{3/2}^3/M_P\ll m_{3/2}^2$)
and thus the effects of uplifting can be neglected.
To compute the scalar masses 
 and A-terms, the key piece of information required is the modular
dependence of the kinetic terms $\tilde{K}_{\alpha}$. In the dilute flux limit this can be derived
by relating the modular scaling of $\tilde{K}_{\alpha}$ to that of the physical
Yukawa couplings $\hat{Y}_{\alpha \beta \gamma}$ through the relation
\be
\hat{Y}_{\alpha \beta \gamma} = \frac{e^{\hat{K}/2} Y_{\alpha \beta \gamma}}
{(\tilde{K}_{\alpha} \tilde{K}_{\beta} \tilde{K}_{\gamma})^{\half}},
\ee
and the fact that the $T$ moduli cannot appear perturbatively in $Y_{\alpha \beta \gamma}$ due to the
combination of holomorphy and the Peccei-Quinn (PQ) shift symmetry.
Ref.~\cite{hepth0609180} used this reasoning to derive
\be
\label{mattermetricc}
\tilde{K}_{\alpha} \sim \frac{\tau_s^{1/3}}{\mc{V}^{2/3}} k_{\alpha}(\phi),
\ee
where $\phi$ refers to the complex structure moduli.
In the dilute flux limit this leads to scalar masses and A-terms given by
\bea
\label{scalarAdil}
m_{\alpha} & = & \frac{1}{\sqrt{3}} \frac{F^s}{2 \tau_s} = \frac{M}{\sqrt{3}}, \nonumber \\
A_{\alpha \beta \gamma} & = & -\frac{F^s}{2 \tau_s} = -M. \\
\label{BTerm}
B & = & -\frac{4 M}{3}.
\eea

The superpotential $\mu$-term can be shown to vanish due to scaling
arguments
\cite{hepth0609180}: in the limit that $\mc{V} \to \infty$,
any non-zero superpotential $\mu$-term would generate a mass for the Higgs field
above the string scale.
A non fine-tuned (i.e.\ weak-scale)
$\mu$ term can be generated through the Giudice-Masiero
mechanism \cite{giudice}, and as such is guaranteed to be of a similar order
to the other soft terms. In fact it turns out to be difficult to satisfy the B-term constraint (\ref{BTerm}) when running to low energies.
However, the Higgs sector is by far the least understood sector of the Standard Model.
As such it may be non-minimal, or the B-term may receive an extra contribution
through the vacuum expectation value of a gauge-invariant scalar,
$\alpha \langle N \rangle H_1 H_2$. In this paper we therefore follow normal phenomenological practice by trading $B$ for $\tan \beta$
and treating $\tan \beta$ as a free parameter.

However, in addition to modifying the gauge couplings, the magnetic
fluxes will also modify the kinetic terms and their dependence on $\tau_s$.
This affects soft breaking terms except in the limit of
large cycle volume and dilute fluxes, where flux dependence should disappear.
While there do not exist explicit formulae for Calabi-Yau compactifications,
for compactifications on toroidal orientifolds such behaviour has been analysed
using string scattering techniques.
A typical result is (see \cite{hepth0404134} and section 4.4 of \cite{hepth0610327})
\be
\mc{K}_{C \bar{C}} = (S + \bar{S})^{-\alpha} \prod_{j=1}^3 (T^j + \bar{T}^j)^{\beta + \gamma \frac{\phi^j}{\pi}}
(U^j + \bar{U}^j)^{-\frac{\phi^j}{\pi}} \sqrt{\frac{\Gamma(\phi^j / \pi)}{\Gamma(1 - \phi^j / \pi)}},
\ee
where $U_j, S$ are the complex structure moduli and dilaton fields
respectively, $\alpha$, $\beta$, $\gamma$ are constants and $\phi^j = \arctan \left( \frac{f^j}{t_2^j} \right)$
is the relative angle between the branes, with $f^j$ a flux quantum number.
We can see that the $T$-dependence involves the angles $\phi_i$. However, in the limit that $T \to \infty$, $\phi_i \to 0$
and the flux-dependence disappears. We also see that fields with different $\phi_{i}$, corresponding to different
gauge charges, experience a different T-dependence.
While toroidal examples are not the case of direct interest, they are important because they
 do allow an explicit computation of flux effects and one can explicitly see how the effects of
fluxes become less important at large cycle volume.

The large-volume models rely on a Calabi-Yau geometry and there do not exist any direct formulae for the effects of magnetic
fluxes on the matter metrics. Nonetheless, the fluxes will affect the soft terms and must be taken into account.
To model the flux corrections, we shall use the simple ansatz
\be
\label{fluxK}
\tilde{K}_{\alpha} = \frac{(\tau_s + \epsilon_{\alpha}(F))^{1/3}}{\mc{V}^{2/3}}.
\ee
$\epsilon_\alpha$ is used to parametrise the flux effects, and in the dilute flux limit
$\epsilon_\alpha \ll \tau_s$. While the form of (\ref{fluxK}) is
simpler than will actually occur, it satisfies the basic requirement that the flux contribution will vanish in the limit
that the cycle size goes to infinity and the fluxes dilute away. Using (\ref{fluxK}) we can then compute
\bea
\label{fluxsoftterms}
m_{\alpha} & = & \frac{1}{\sqrt{3}} \frac{F^s}{2 (\tau_s +
  \epsilon_{\alpha}(F))}, \nonumber \\
A_{\alpha \beta \gamma} & = & -\frac{1}{\sqrt{3}} \left( m_{\alpha} + m_{\beta} + m_{\gamma} \right).
\eea
In the limit that $\epsilon_{\alpha} \to 0$, this recovers the dilute-flux expressions (\ref{scalarAdil}).

\subsection{Flavour and CP issues}
\label{flavoursubsec}

The formulae (\ref{SoftMassDiagFormula}) to (\ref{BTermFormula})
used above in computing the soft terms are a restricted form applicable for flavour-diagonal soft masses.
This requires some explanation, as `generic'
gravity-mediated models give flavour non-universal soft terms and corresponding problems with flavour-changing neutral currents.
The problem with this `generic' expectation is that it is essentially a
naive argument using only effective field theory, which does not take into
account the actual structures that arise in string compactifications, which violate the genericity assumptions.

It was argued in \cite{hepth0610129} that
for the large-volume models, and more generally for models arising from IIB flux compactifications,
there is a clean understanding of why supersymmetry breaking should give universal soft
terms, at least at leading order. We briefly summarise the argument.
The ability to distinguish flavours and consider flavour mixing comes from the structure of the Yukawa couplings.
The Yukawa couplings are generated from the superpotential,
and as such can only depend on the dilaton and complex structure moduli. The combination of
holomorphy and the PQ shift symmetry $\hbox{Im}(T) \to \hbox{Im}(T) + \epsilon$ implies that the K\"ahler moduli cannot make
any perturbative appearance in the superpotential:
since
flavour is generated in the superpotential,
the interactions of the $T$-fields are flavour-blind. The physical Yukawa couplings also depend on the K\"ahler metrics; the scaling
arguments entering (\ref{mattermetricc}) however guarantee that different flavours have the same scaling with the $T$-moduli.

However, in IIB flux models it is
the $T$-fields that have non-vanishing F-terms and break
supersymmetry. The K\"ahler potential (\ref{KahlerPot})
also has a block-diagonal
structure,
\be
K_{\Phi \bar{\Phi}} = \left(
\begin{array}{ccc} K_{S \bar{S}} & 0 & 0 \\
0  & K_{U \bar{U}} & 0 \\
0 & 0 & K_{T \bar{T}}
\end{array} \right) + \mc{O} \left( \frac{1}{\mc{V}} \right).
\ee
Mixing between the $T$ and $S$,$U$-fields is volume-suppressed and tiny. Thus the fields that break supersymmetry - the
K\"ahler moduli - and the fields that give flavour - the $S$ and $U$
fields - are decoupled, to leading order, 
 and supersymmetry breaking generates
flavour universal soft terms.

Large CP-violating phases are likewise not a problem. From the structure of the soft
terms (\ref{CNGM}), (\ref{SoftMassDiagFormula}) and
(\ref{ATermFormula}),
we
see that the gaugino and A-term phases are all
inherited from the small modulus F-term $F^s$. They are
thus universal and do not generate large CP violating phases that would be in
conflict with observations.

\section{Spectra}
\label{spectra}

\subsection{Generation}

The low-energy mass spectrum is determined by evolving the soft terms
from the high scale to the TeV scale. Ambiguities in the high-scale
soft terms will translate into ambiguities in the physical spectrum
and observables. In ref.~\cite{hepth0610129}, a basic phenomenological
analysis was carried out using the soft terms in the dilute-flux
limit, (\ref{DiluteFluxGauginoMasses}) and (\ref{scalarAdil}), which were run down to produce TeV mass spectra.
However, as emphasised above, the contribution of magnetic fluxes automatically
introduces a theoretical uncertainty in the high-scale soft
parameters. We want to understand how this uncertainty manifests
itself in the possible low energy spectra.

We make the assumption that the cycles are sufficiently large that
the effect of the fluxes is as a perturbation on the dilute-flux
results. In general the fluxes are unknown, but in one instance they
can be taken `from data'. In the dilute-flux limit, from
(\ref{dilutef}) the non-abelian gauge couplings would be exactly
universal at the high scale $m_s \sim 10^{11} \hbox{~GeV}$. From
(\ref{f3}) and (\ref{f2}), it follows that the fluxes are directly
responsible for the difference of $SU(2)$ and $SU(3)$ gauge
couplings. The magnitude of the flux perturbation can then be
estimated from this ratio.

Assuming an MSSM spectrum with no exotic matter, the ratio of
high-scale gauge couplings is
\be \label{g3g2}
\frac{g_3^2}{g_2^2}\Bigg|_{10^{11}{\rm GeV}} \approx 1.37. \ee We regard
both $g_3^2$ and $g_2^2$ as fluctuations
around a common value. In
determining the soft terms that we run down to generate low energy
spectra, we use the following strategy. We first specify a high
scale value for the gluino mass, $M_3$. The relations
(\ref{m3m2ratio}) and (\ref{g3g2}) then fix the high-scale wino
mass, \be M_2 \Big\vert_{m_s = 10^{11} \hbox{~GeV}} \approx \frac{1}{1.37}
M_3 \Big\vert_{m_s = 10^{11} \hbox{~GeV}}. \ee The gaugino masses are
flux-induced perturbations about a base value $\frac{F^s}{2 \tau_s}$. We use the
central value of $M_2$ and $M_3$ to estimate this,
\be
\frac{F_s}{2\tau_s} \approx M_c = \frac{M_2 + M_3}{2}.
\ee
We generate the
remaining soft masses as fluctuations about $M_c$. For example,
\bea
\label{genSoftTerms}
M_1 & = & M_c (1 \pm \epsilon_1), \nonumber \\
m_a & = & \frac{M_c}{\sqrt{3}} (1 \pm \epsilon_a).
\eea
Here $m_a$ stands for the soft breaking mass for scalars.
The perturbation parameter $\epsilon_a$ differs for each type of field, but is
 assumed to be the same across generations.\footnote{This is required from considerations of flavour physics.
 The theoretical justification for this is that the flux magnitudes are dual to brane intersection angles $\theta_i$,
 and fields of different flavour but with the same gauge charges see the same angles $\theta_i$. It would be useful to further
 examine this question within explicit models.}

With the ansatz (\ref{fluxK}), the A-terms are given in terms of
the scalar masses by (\ref{fluxsoftterms}).
\be
\label{aterm}
A_{\alpha \beta \gamma} = - \frac{1}{\sqrt{3}} (m_{\alpha} +
m_{\beta} + m_{\gamma}).
\ee

 We first generate a set of high-scale soft terms according to the relations
(\ref{genSoftTerms}).
For each spectrum, the $\epsilon_a$ were randomly
generated within a domain $0 < \epsilon_a < \epsilon_0$ with
constant probability density.
We initially take $\epsilon_0 = 0.2$, but also investigate the choice $\epsilon_0 = 0.4$.
In generating spectra, as stated above we do not impose the high-scale value for the B-term (\ref{BTerm})
and instead treat $\tan \beta$ as a free parameter, which we allow to lie in the region
$5 < \tan \beta < 40$.
We repeatedly generate many random spectra in this manner. We then remove any
spectra which fail experimental constraints, although all points pass direct
sparticle search limits due to the heavy SUSY breaking scale set.

Using the program \verb+SOFTSUSY2.0+ \cite{hepph0104145} the 2-loop
renormalisation group equations
(RGEs) for the MSSM are solved numerically to obtain a particle
spectrum at the weak scale.
The values of the Standard Model input parameters used for our
computations are $m_t = 171.4$ GeV
\cite{hepex0608032}, $m_b (m_b)^{\overline{MS}} =4.25$ GeV,
${\alpha}_s (M_Z)^{\overline{MS}} = 0.1187$, $\alpha^{-1}
(M_Z)^{\overline{MS}} = 127.918$, $M_Z = 91.1187$ GeV \cite{pdg}.
Every particle spectrum generated
must be regarded as equally consistent with the
large-volume scenario we study. In determining what counts as an acceptable
spectrum, we impose experimental constraints on
the magnitude of $BR(b\to s\gamma)$, the anomalous magnetic moment of the muon $(g-2)_\mu$ and
the dark matter relic density $\Omega h^2.$ We require spectra to
generate values for these within $2 \sigma$ of
the experimental results.

The average measurement of the $b\to s\gamma$ branching ratio was
obtained from \cite{heavyflavour}: $BR(b\to s\gamma) = (3.55\pm
0.26) \ti 10^{-4}.$ For an estimate of the theoretical uncertainty we used the result of
\cite{hepph0410155} $0.30\ti 10^{-4}$; adding the two errors in
quadrature, we obtain the $1\sigma$ bound, $BR(b\to s\gamma) =
(3.55\pm 0.40)\times 10^{-4}.$

The anomalous muon magnetic moment $a_\mu = (g_\mu-2)/2$ represents
the largest theory/experiment discrepancy in precision electroweak
physics. The average experimental value of $a_\mu$ is
$116592080(63)\times 10^{-11}$ \cite{hepex0401008}, with an error
that is statistics dominated. The dominant uncertainties in the
Standard Model computation of $a_\mu$ arise from hadronic
light-by-light scattering and vacuum polarisation diagrams - for
recent reviews see \cite{hepph0703049, hepph0703125}. The evaluation
of the vacuum polarisation diagrams is carried out using
experimental data from $e^+ e^-\to {\ \rm hadrons}$.
The Standard Model
result given in \cite{hepph0703125} is $a_{\mu}^{the} =
116591785(61) \times 10^{-11}$, giving a 3.2 $\sigma$ discrepancy
$$
\delta a_\mu = (287 \pm 91) \ti 10^{-11}.
$$
It is important to note that $\delta a_{\mu}$ usually has the same
sign as $\mu$, so a supersymmetric explanation of the $g_\mu - 2$
discrepancy prefers $\mu > 0$.

We also impose bounds on the Higgs mass obtained by the LEP2
collaborations \cite{hepex0306033}. The lower bound is $114.4$ GeV
at the 95\% CL.
The theoretical computation of the Higgs mass in supersymmetric
scenarios is subject to an estimated error of $3$ GeV,
and so we impose $m_h>111$ GeV on the result obtained from
\verb+SOFTSUSY+.

The $2 \sigma$ WMAP \cite{WMAP} constraint on the relic density of dark matter
particles is
\begin{equation}
\label{omh2}
0.085 < \Omega h^2 < 0.125.
\end{equation}
Assuming a thermal relic abundance, we compute the neutralino contribution to dark matter using
\verb+micrOmegas1.3+  \cite{hepph0405253}. The lower bound in
(\ref{omh2}) is only applicable if we require that the dark matter is solely composed of
neutralinos. Since there may be other dark matter constituents, such
as axions or for the large volume models the volume modulus \footnote{The volume modulus is light and has the potential to overclose the universe.  Its abundance must therefore be diluted.  We refer interested readers to \cite{hepph07053460} for a detailed discussion on this issue, 
as well as other astrophysical and cosmological implications.}, when considering spectra we only impose the upper bound on $\Omega h^2$.

In determining the collider phenomenology
the most important feature of any SUSY spectrum is the overall
scale, in particular the scale of the squarks and gluinos
whose pair production initiates the majority of supersymmetric
events at a hadron collider. This has a large but mostly trivial effect on the observables,
primarily through an overall rescaling of the number and energy scale of SUSY
events: the lighter the spectrum, the more events that are generated.
In studying the different spectra
produced by the large-volume models, and how the high-scale
flux uncertainty translates into a low-scale spectrum uncertainty,
our interest is not so much in the overall scale of the spectrum as in its
structure. We therefore
take a fixed value $M_3 = 500 \hbox{~GeV}$ at the high scale, which
corresponds to a physical gluino mass $m_{\tilde{g}} \sim 900
\hbox{~GeV}$. Assuming supersymmetry is discovered at the LHC, the overall
production scale could be constrained by a variable such as
$M_{eff}$~\cite{meff}.
In section 4 we will consider the effects of varying the
overall scale.

\subsection{Features}

The particle masses of 200 spectra passing the 2$\sigma$ experimental
constraints and randomly generated with $M_3 = 500
\hbox{~GeV}$ at $m_s = 10^{11} \hbox{~GeV}$ and $20 \%$ variation parameters
$\epsilon_\alpha$, are plotted in figure \ref{massesM3is500} \footnote{The dilute flux spectrum shown is chosen to give a similar sparticle production scale compared to the other spectra shown in the figure.  Another useful comparison would be a dilute flux spectrum with $M_3 \approx 430 \hbox{~GeV}$, which corresponds to the base value $M_c$ used for generation of the large volume models.  This spectrum is not expected to lie at the centre of the large volume model spectra because of the different gaugino masses defined at the string scale, which in turn implies different RG evolution.\label{dilutefluxcomment}}. From figure \ref{massesM3is500}
we can see the extent to which uncertainties at the high scale translate into uncertainties at the low scale.
It should be noted that because the fluctuations are assumed to be the same for each generation,
but to vary independently for each type of field, there are correlations among fields of different generations.
\FIGURE{\makebox[15cm]{\epsfxsize=15cm \epsfysize=10cm \epsfbox{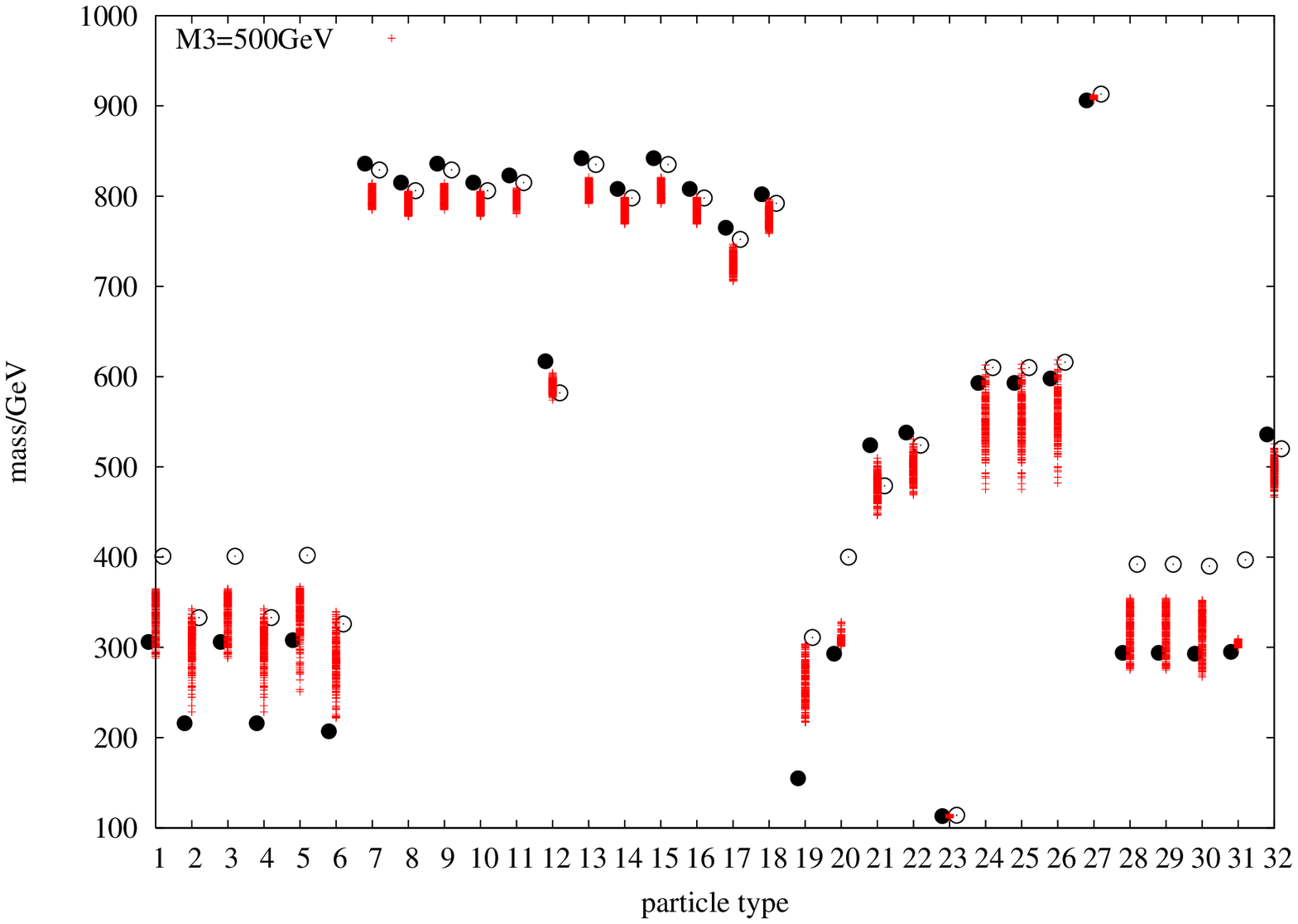}} \caption{Particle spectra with the gluino mass
normalised to $\sim 900 \hbox{~GeV}$ and $20\%$ high scale
uncertainties in other parameters.
For comparative purposes, an
SPS1a spectrum with the same value
of the gluino mass is also shown using black dots. The hollow circles show a spectrum coming from soft terms
in the dilute flux limit. Particles 1-6 are the
left and right handed sleptons, particles 7-18 are the squarks, 19 is
the lightest supersymmetric particle (LSP)
 $\tilde{\chi}^0_1$, 20-22 are the remaining $\tilde{\chi}_i^0$, 23-26 the
Higgs particles, 27 the gluino, 28-30 the sneutrinos, 31 and 32 are
the charginos $\tilde{\chi}_{1,2}^\pm.$\label{massesM3is500}}}

All of these spectra satisfy the above observational
constraints. The $\tan\beta$ parameter was allowed to be in the range
$5 < \tan \beta < 40$, but all of the points that pass the constraints have
$\tan\beta < 23.$
The generic features of the spectra are
\begin{enumerate}
\item The first-generation squarks are heavy, with masses
around 800GeV. The $m_{\tilde{q}_1} : m_{\tilde{g}}$ mass ratio is
well-predicted, with a range from 780:900 to 810:900. 
The lighter stop (particle 12 in the figure)
is at around 600 GeV, while the sbottom (particle 17) has a 720 GeV mass.
\item The sleptons are much lighter at around $300 \hbox{~GeV}$. The fluctuations in the
$m_{\tilde{l}} : m_{\tilde{g}}$ ratio are much larger than for the squark
and inherit their magnitude from the high-scale fluctuations. The stau is the
lightest slepton, but is comparable in mass to the $\tilde{e}$ and $\tilde{\mu}.$
\item
The chargino $\tilde{\chi}_1^{\pm}$ has $m_{\tilde{\chi}_1^{\pm}} = 304
\pm 5 \hbox{~GeV}$ and exhibit very little fluctuation.
It is nearly degenerate with the second neutralino $\tilde{\chi}^0_2$, which tends to be mostly wino.
\item
The LSP tends to be mostly bino, with a mass that can
fluctuate substantially between 200GeV and 300GeV. If the LSP has a mass towards the top end
of this range it can have a substantial wino component.
\item
The charged Higgs fields are intermediate between the squarks and sleptons, with masses
around $500 \hbox{~GeV}$.
\end{enumerate}

The choice of $20 \%$ for the variation parameter $\epsilon_0$ is somewhat arbitrary. We have
also generated spectra in which the variations were only constrained to be within $40 \%$ of the central
value $M_c$. These spectra are shown in figure \ref{masses40percentM3is500} \footnote{See footnote \ref{dilutefluxcomment} for comments on the dilute flux spectrum.}.
\FIGURE{\makebox[15cm]{\epsfxsize=15cm \epsfysize=10cm \epsfbox{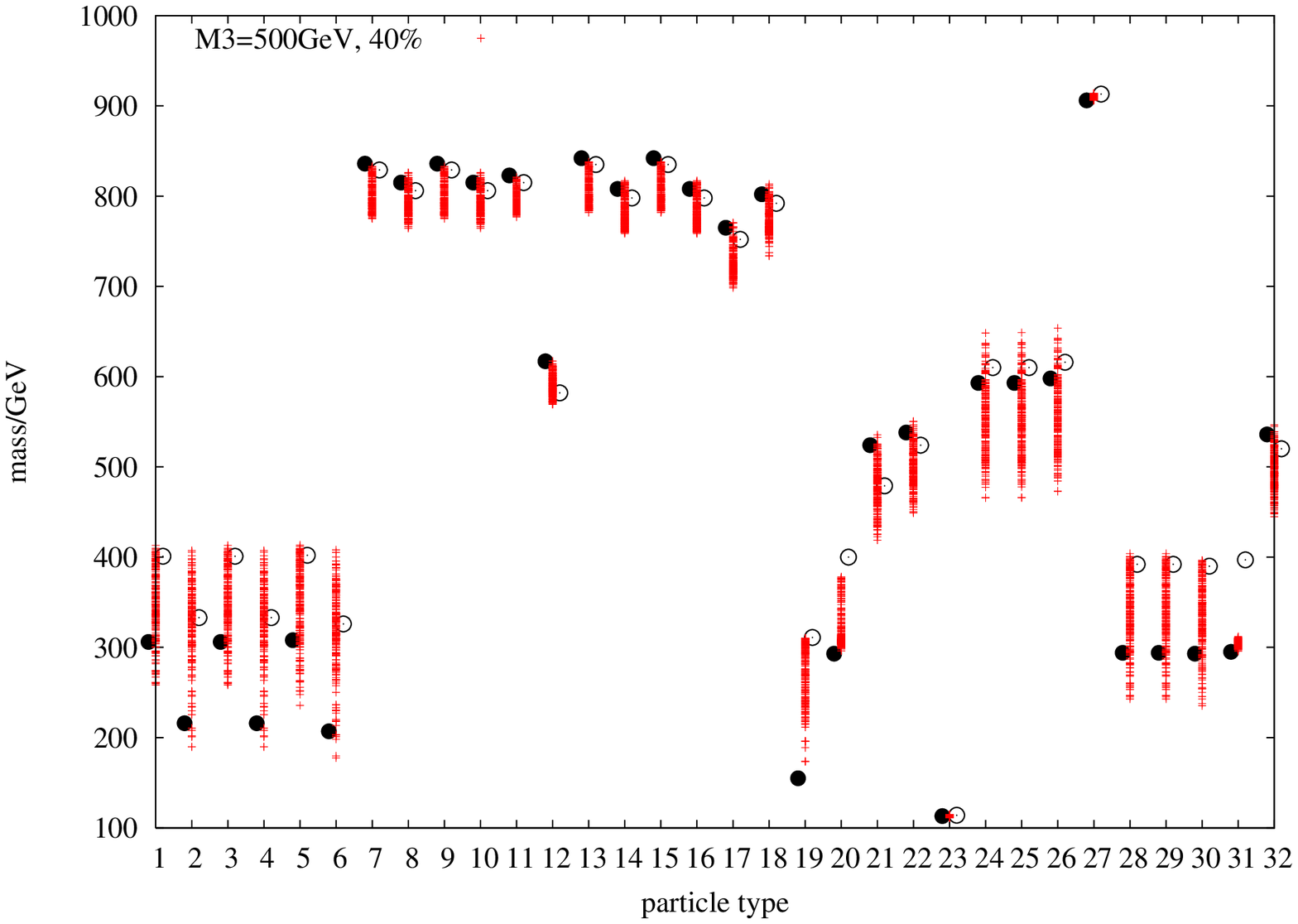}} \caption{Particle spectra with the gluino mass
normalised to $\sim 900 \hbox{~GeV}$ and $40\%$ high scale
uncertainties in other parameters.
For comparative purposes, an
SPS1a spectrum with the same value
of the gluino mass is also shown using black dots.
The hollow circles show a spectrum coming from soft terms
in the dilute flux limit.
Particles 1-6 are the
left and right handed sleptons, particles 7-18 are the squarks, 19 is
the LSP $\tilde{\chi}^0_1$, 20-22 are the remaining $\tilde{\chi}_i^0$, 23-26 the
Higgs particles, 27 the gluino, 28-30 the sneutrinos, 31 and 32
the charginos $\tilde{\chi}_{1,2}^\pm$.\label{masses40percentM3is500}}}

The spectra with $40 \%$ variation in high-scale parameters exhibits the same basic structure as that with
$20 \%$ variation, except that the spread is greater. The variation in squark masses remains much less than that for
weakly interacting sparticles.

The features of the spectrum described above can be explained analytically.
The simplest example is for the ratios of
gaugino masses. The one-loop RGEs for the gauge couplings and gaugino
masses are
\bea
\label{gaugerunning}
\frac{dg_a}{dt} & = & \frac{g_a^3}{16 \pi^2} b_a, \\
\label{gauginorunning}
\frac{dM_a}{dt} & = & \frac{2 g_a^2}{16 \pi^2} b_a M_a.
\eea
It follows from (\ref{gaugerunning}) and (\ref{gauginorunning}) that
at one-loop
\be
\frac{d}{dt} \left( \frac{M_a}{g_a^2} \right) = 0,
\ee
and so
\be
\frac{M_3}{M_2}\Bigg|_{M_Z} = \frac{g_3^2}{g_2^2}\Bigg|_{M_Z} \ti
\left( \frac{M_3 g_2^2}{M_2 g_3^2} \right)\Bigg|_{M_X} = \frac{g_3^2}{g_2^2}\Bigg|_{M_Z},
\ee
where we have used the high-scale gaugino mass expressions
(\ref{PhysicalGauginoMasses}).
For the bino mass, we have
\be
\frac{M_2}{M_1}\Bigg|_{M_Z} = \frac{g_2^2}{g_1^2}\Bigg|_{M_Z} \ti \left(
\frac{M_2 g_1^2}{M_1 g_2^2} \right)\Bigg|_{M_X} = \frac{g_2^2}{k_Y g_1^2}\Bigg|_{M_Z}.
\ee
The bino mass cannot then be taken directly from data as it
depends on the factor $k_Y$ which enters the $U(1)$ normalisation.
The fluctuations in the bino mass at the weak scale are inherited from its
high-scale fluctuations
and are large. The bino tends to be heavier than in mSUGRA models. This can be understood from the
leading order universality of the gaugino masses at the intermediate scale, which holds up to the effects of magnetic fluxes and
tends to compress the gaugino mass spectrum.\footnote{A similar compression is observed in mirage mediation models, where
the gaugino masses are exactly universal at the intermediate scale.}
It may be possible to achieve an $M_1:M_2$ ratio comparable to that of mSUGRA, but as seen in figure \ref{masses40percentM3is500}
it would be non-generic and
difficult to achieve while still treating the magnetic fluxes as perturbations.

To explain the structure of scalar masses and the magnitude of their
fluctuations, we can use the approximate analytic solution for the
soft parameters that is obtained by one-loop RGE running to the low scale \cite{hepph9709356}:
\begin{eqnarray}
\label{analyticscalars}
m_{\tilde{d}_L}^2 \Big\vert_{Q} &=& m_{\tilde{d}_L}^2\Big\vert_{Q_0} + K_3 + K_2 + {1\over{36}} K_1 +\Delta_{\tilde{d}_L}, \nonumber \\
m_{\tilde{u}_L}^2 \Big\vert_{Q} &=& m_{\tilde{u}_L}^2\Big\vert_{Q_0} + K_3 + K_2 +  {1\over{36}} K_1 + \Delta_{\tilde{u}_L}, \nonumber \\
m_{\tilde{u}_R}^2 \Big\vert_{Q} &=& m_{\tilde{u}_R}^2\Big\vert_{Q_0} + K_3 + {4\over9} K_1 + \Delta_{\tilde{u}_R}, \nonumber \\
m_{\tilde{d}_R}^2 \Big\vert_{Q} &=& m_{\tilde{d}_R}^2\Big\vert_{Q_0} + K_3 + {1\over9} K_1 + \Delta_{\tilde{d}_R},\nonumber \\
m_{\tilde{e}_L}^2 \Big\vert_{Q} &=& m_{\tilde{e}_L}^2\Big\vert_{Q_0} + K_2 + {1\over4}K_1 + \Delta_{\tilde{e}_L},\nonumber \\
m_{\tilde{e}_R}^2 \Big\vert_{Q} &=& m_{\tilde{e}_R}^2\Big\vert_{Q_0} + K_1 + \Delta_{\tilde{e}_R}.
\end{eqnarray}
Here $K_i$ is determined by the RGE running of the gaugino mass
$M_i$, and is given by
\bea
\label{kvalues}
K_a(Q) & = & \left\{ \begin{array}{c} 3/5 \\ 3/4 \\ 4/3 \end{array} \right\}
\times \frac{1}{2 \pi^2} \int_{\ln Q}^{\ln Q_0} dt \, g_a^2(t) \vert M_a(t) \vert^2 (a = 1,2,3) \nonumber \\
& = & \left\{ \begin{array}{c} 6/33 \\ 3/2 \\ -8/9 \end{array} \right\} \left(
M_a^2 \Big\vert_{Q_0} - M_a^2 \Big\vert_{Q} \right).
\eea
where $Q_0$ is the high scale and $Q$ is the scale at which the squark and slepton masses are
evaluated. The $\Delta$ contributions come from the D-terms and are
expected to be small, being proportional to $m_Z^2$.

For the large-volume models we
numerically obtain $K_3\approx 1.92
\, M_3^2 (M_X)$ with $M_3 (M_X) = 500$ GeV, $M_X = 10^{11}$ GeV. For comparison,
an mSUGRA point on the SPS1a slope has $K_3 \approx 3.84 \, M_3^2 (M_X)$ with $M_3
(M_X)=500$ GeV, $M_X\sim 10^{16}$ GeV.

The effect of (\ref{analyticscalars}) and (\ref{kvalues}) is that as we run to
low energies the gluino mass comes to give the dominant contribution to the squark masses.
This is due to the large contribution of the $K_3$ factor
in equations (\ref{analyticscalars}). For small $m_{\tilde{q}}(M_X)$,
$m_{\tilde{q}}(M_{SUSY})$ is mainly determined by the gluino mass
and is relatively insensitive to the initial
value of $m_{\tilde{q}}(M_X)$. Here $M_{SUSY}$ is the scale of the SUSY breaking
spectrum, $\sim \mc O (1)$ TeV.
This explains why the fluctuations in
$m_{\tilde{q}}(M_{SUSY})$ are small, since the gluino mass is fixed at
tree-level by $M_3=500$ GeV.
The slepton evolution is however driven only by the wino and bino,
corresponding to the weakly coupled $SU(2)$ and
$U(1)$ gauge groups.
RGE effects from the winos and binos are less pronounced than those of
the gluino, and thus
$m_{\tilde{l}}(M_X)$
gives the dominant contribution to
$m_{\tilde{l}}(M_{SUSY})$. The physical slepton masses inherit the high-scale
fluctuations of $m_{\tilde{l}}(M_X)$ and
this produces the large slepton mass fluctuations observed in
figures \ref{massesM3is500} and \ref{masses40percentM3is500}.

For the spectra of figures \ref{massesM3is500} and \ref{masses40percentM3is500},
the stau is lighter than the other sleptons, but the stau-slepton mass splitting is not as pronounced as
mSUGRA models with the same $\tan \beta$. This is because there is less room for the stau to evolve when running from
the intermediate scale rather than the GUT scale,
and because the larger value of $M_1$ tends to compensate for the effect of the
large Yukawa coupling in the stau RGEs.

It is promising that even allowing for large $40 \%$ variations in the high-scale soft terms does not
alter the overall structure of the low-energy spectrum. We now describe how the spectra of
figure \ref{massesM3is500} and \ref{masses40percentM3is500}
can be distinguished from mSUGRA or mirage mediation models.

\subsection{Discrimination from mSUGRA models}

The spectra that appear in figures \ref{massesM3is500} and \ref{masses40percentM3is500}
have properties of `typical' gravity-mediated scenarios and are
broadly similar to those arising from mSUGRA models.
However, there are several important differences which enable these to be discriminated.

The most obvious is the gaugino mass ratio. In mSUGRA the ratios of all three gaugino masses
are set by the gauge couplings, with low-scale values
\be
\label{msugragauginoratio}
\left( M_3 : M_2 : M_1 \right)\Big|_{M_Z} = \left( g_3^2 : g_2^2 : g_1^2 \right) \Big|_{M_Z} \sim 6 : 2 : 1.
\ee
Here $g_1^2 = \frac{5}{3}g_Y^2$ with the GUT normalisation of $U(1)_Y$.
This ratio follows from gaugino mass universality at the GUT scale.
From the spectrum of figure \ref{massesM3is500}, we see that for the large-volume models we have
\be
\label{largevolgauginoratio}
\left( M_3 : M_2 : M_1 \right)\Big|_{M_Z} = \left(g_3^2 : g_2^2 : k_Y g_Y^2 \right)\Big|_{M_Z} \sim 6 : 2 : (1.5 \to 2).
\ee
(\ref{largevolgauginoratio}) thus
gives both a distinct prediction for the $m_{\tilde{g}} : m_{\tilde{W}}$ mass ratio
together with the expectancy that the $m_{\tilde{W}} : m_{\tilde{B}}$ ratio will differ from that in
mSUGRA. We can understand the larger value of the bino mass relative to mSUGRA from the fact that gaugino masses are
approximately universal at the intermediate scale, up to the effects of magnetic fluxes.
To achieve the mSUGRA gaugino mass ratios in the large volume models requires a very large
fluctuation in the high-scale bino mass away from the value expected in the dilute-flux approximation.
 This is non-generic and difficult to achieve while still regarding the magnetic fluxes
as perturbations. An observation of the gaugino mass ratios
(\ref{msugragauginoratio}) would thus disfavour the large-volume scenario.

The use of gaugino mass ratios as observables has one distinct advantage, recently emphasised in ref.~\cite{hepph0702146}. The matter content enters both the gaugino and gauge coupling RGEs in the same way (through $b_a$)
and so the one-loop derivation of the ratios (\ref{msugragauginoratio}) and
(\ref{largevolgauginoratio}) is independent of
any extra charged matter that may be present beyond the MSSM. String constructions generically contain extra vector-like matter
that will affect the running of the gauge couplings. The point is that such matter will equally affect the
running of the gaugino masses and gauge couplings,
and so the low-energy ratios (\ref{msugragauginoratio}) and (\ref{largevolgauginoratio})
will be unaltered. In this respect these represent predictions that do not rely on the details of the matter
spectrum. Another advantage of the gaugino mass ratios (\ref{largevolgauginoratio}) is that the derivation of these soft terms
shows that these ratios persist even beyond the dilute flux approximation, indeed up to the point where non-perturbative
string corrections in $e^{-T_s}$ would become important in evaluating the gauge kinetic functions.

There is a caveat to be added concerning the above relationship of gaugino masses (\ref{largevolgauginoratio}).
Two factors enter into the determination of the ratio $M_1 : M_2$. The first is requiring that the high-scale perturbations remain
in the dilute-flux regime. As an extreme example, within the 20\% and 40 \% variations above it is never possible to get
$M_1 : M_2$ as $1:10$; such a result requires high-scale soft terms that violate the dilute-flux assumption.
The second factor is that in determining the spectra that count as allowed in figures 
\ref{massesM3is500} and \ref{masses40percentM3is500} above, the upper bound on the
WMAP relic density constraint ($\Omega h^2 < 0.11$) has been imposed.
As the bino mass enters crucially into the evaluation of the relic density, this imposes a further selection cut on
allowed values for the $M_1: M_2$ ratio, in conjunction with the slepton masses.

The first constraint (remaining with the dilute flux assumption) prefers an $M_1 : M_2$ ratio larger than found in mSUGRA,
i.e. greater than $1:2$. The second constraint strengthens this preference. For example in the 40\% diagram, the spectra with the
lowest values of $M_1$ are also seen to have low values for the slepton masses (this allows e.g stau coannihilation to reduce the
relic density to the WMAP allowed region). This requires a large downward
flux-induced perturbation of the slepton masses in correlation with that of the bino mass, which is disfavoured.

The spectra shown in figures \ref{massesM3is500} and \ref{masses40percentM3is500}, and the resulting gaugino mass ratio of \ref{largevolgauginoratio}, are a convolution of the above two effects. Actually, it is expected that this discussion will need further modification. The reason is that the use of
the thermal relic abundance computation relies on the assumption that cosmology is thermal from a temperature of $T \sim \mc{O}(10 \hbox{GeV})$
down. Such reheating temperatures are in general hard to achieve within string-derived theories due to the cosmological moduli problem.
In the large-volume models, there exists a light gravitationally coupled $\mc{O}(\rm{MeV})$ volume modulus. Such a field has a lifescale longer
than the universe and must be diluted for it not to spoil nucleosynthesis.\footnote{The cosmological properties of the large volume models
are discussed at more length in \cite{hepph07053460}.} Some dilution mechanism, such as thermal inflation, is therefore necessary to dilute the
modulus. The presence of such a dilution mechanism alters the late-time cosmology. The proper computation of the relic abundance
should start with the reheating temperature that applies after the dilution mechanism has been in operation. The dilution mechanism
may itself lead to misalignment of the light modulus and thus require a reheating temperature much less than $\mc{O}(10) \hbox{GeV}$.

The upshot of this discussion is that to avoid the moduli problem the reheating temperature may end up being significantly less
than assumed in the thermal computation of the relic abundance, and this will tend to modify the spectra and gaugino mass ratios above.
A lower reheating temperature will imply the need for more efficient particle annihilation. In general this will tend to prefer greater
degeneracies in the sparticle spectrum (e.g. to allow efficient coannihilation channels), but the detailed effects of a lower reheating
temperature are beyond the scope of this paper.

We note the spectra (\ref{msugragauginoratio}) and (\ref{largevolgauginoratio}) could
potentially be distinguished within the first year of LHC running; for example a
di-lepton edge measurement $m_{\tilde{\chi}_2^0} - m_{\tilde{\chi}_1^0} = 50 \hbox{~GeV}$ together with an estimated
SUSY production scale of $m_{SUSY} \sim 1000 \pm 300 \hbox{~GeV}$ is consistent with the second gaugino spectrum
and not with the first.

A more subtle distinction can be made through the relative values of the scalar masses. For example, the evolution
of the first-generation squarks is largely driven by the gluino and the RGEs for
these squarks exhibit focusing behaviour. If we
neglect all but the strong dynamics, we have a solution of the one-loop
RGEs~\cite{allanach}
\begin{equation}
\frac{m_{\tilde q}^2(\mu)}{M_3^2(\mu)}
= \frac{8}{9} (1 - r^2) + r^2 \frac{m_{\tilde q}^2(M_X)}{M_3^2(M_X)}
\qquad
\mbox{where} \qquad
r = \frac{g_3^2 (M_X)}{g_3^2(\mu)}. \label{focusing}
\end{equation}
As the renormalisation scale $\mu$ is lowered, an infra-red stable fixed point
of $r \rightarrow 0$ is approached, i.e.
 $\left( \frac{m_{\tilde q}^2}{M_3^2} \right)
 \rightarrow \frac{8}{9}$. However, how close one gets to this infra-red fixed
 point limit for $\mu=M_{SUSY}$
 depends upon the starting scale of evolution: for $M_X \sim
 10^{16}$ GeV (i.e. for the mSUGRA case), $r = 0.34$, whereas for the large
 volume models, $M_X \sim  10^{11}$  GeV, $r=0.45$.
Equation~(\ref{focusing}) shows that it is generally hard to obtain $m_{\tilde{q}} \ll m_{\tilde{g}}$; if we start
with $m_{\tilde{q}} \sim m_{\tilde{g}}$ then this remains the case, and even
if initially $m_{\tilde{q}} \ll m_{\tilde{g}}$ then the squark will rapidly run up to obtain a mass
$m_{\tilde{q}} \sim m_{\tilde{g}}$.
The point is that by starting the evolution at the intermediate scale,
there is simply less time for the $m_{\tilde{q}} : m_{\tilde{g}}$ ratio to evolve
compared to starting at the GUT scale. If the squark masses start below the gluino mass,
as holds in our case, they have less time to evolve and so tend to be lighter than for an evolution commencing
at $M_{GUT}$: the focusing behaviour of equations (\ref{focusing}) is less efficient.
The ratio $\left( \frac{M_3^2}{m_{\tilde{q}}^2} \right)$ for an intermediate scale model
will then always be less than for a corresponding GUT-scale model with the same soft terms.

This is manifest in our spectrum. We obtain a ratio $m_{\tilde{q}_1} : M_{\tilde{g}} \sim 800:900$, and which can be
significantly smaller (down to $m_{\tilde{q}_1} : M_{\tilde{g}} \sim 770:900$ with the $40 \%$ variation).
A similar choice of high-scale soft terms in mSUGRA gives $m_{\tilde{q}_1} : M_{\tilde{g}} \sim 850:900$, illustrating the
greater running of the squark masses starting from the GUT scale. In the mSUGRA framework, lighter scalar masses at the
GUT scale will reduce the $m_{\tilde{q}_1} : M_{\tilde{g}}$ ratio at the low scale. However the lighter scalar masses also give rise
to lighter sleptons. This can be seen by examining the SPS1a spectrum in figures \ref{massesM3is500} and \ref{masses40percentM3is500}.
While the squark masses are now relatively close to those occurring for the large volume models,
at $m_{\tilde{q}_1} : M_{\tilde{g}} \sim 830:900$,
the sleptons are significantly lighter.
It is thus not possible to fit the scalar spectrum of figures \ref{massesM3is500} and \ref{masses40percentM3is500}
with mSUGRA models unless the assumption of universal scalar mass at
the GUT scale is relaxed: with mSUGRA either the squarks are heavier or the sleptons lighter than for the large volume models.

This illustrates a general feature of the large-volume spectrum, which is that it is more compressed than those appearing in mSUGRA.
This follows primarily from the fact that
the soft term running starts at the intermediate scale rather than the GUT scale. The masses
therefore have less time to separate compared to mSUGRA models, and thus the spectrum is more bunched.

Clearly, the precise $m_{\tilde{q}}:m_{\tilde{g}}$ and $m_{\tilde{l}}:m_{\tilde{g}}$ ratios are not quantities
that can be rapidly measured at the LHC and would
require years of high-luminosity running. However, while difficult to
measure, these ratios are very interesting
because they offer the possibility of estimating the scale at which the soft terms have been defined, and thus even
the possibility of indirectly measuring the string scale.

\subsection{Discrimination from Mirage Mediation Models}

A scenario that has recently attracted attention is mirage mediation
\cite{hepth0503216, hepph0504036, hepph0504037} - see \cite{hepph0612258, hepph0703024, hepph0703163, hepph0507110} for recent work.
This corresponds to soft terms
arising from supersymmetric KKLT stabilisation, with supersymmetry
broken by an anti-D3 brane. The gravitino mass is determined
by the fluxes and is naturally at the Planck scale. The hierarchy is generated by
fine-tuning the fluxes to reduce the gravitino mass to a TeV. The soft terms arise from a combination of gravity and
anomaly mediation. The soft terms are defined at the GUT scale and exhibit mirage unification at an intermediate scale,
\be
M_{mirage} = \left( \frac{m_{3/2}}{M_P} \right)^{\alpha/2} M_P,
\ee
where $\alpha$ is the ratio of gravity to anomaly mediation and is
usually taken to be $\mc{O}(1)$ (see however the discussion in section
5 of \cite{hepth0610129}).
In terms of this parameter the gaugino mass ratios are given by \cite{hepph0702146}
\be
\label{miragegauginoratio}
\left( M_3 : M_2 : M_1 \right)\Big|_{M_Z} =  (6 - 1.8 \alpha) : (2 + 0.2 \alpha): (1 + 0.66 \alpha).
\ee

The $m_{\tilde{W}}: m_{\tilde{B}}$ mass ratio can be similar to that arising from the large-volume models,
and thus it will  not be possible to distinguish mirage mediation and large volume models
from the $m_{\tilde{W}} : m_{\tilde{B}}$ mass ratio. However the $m_{\tilde{g}}:m_{\tilde{W}}$
ratio is substantially smaller for mirage mediation than for the large volume models, and thus here discrimination is
possible. For example, assuming $\alpha \sim 1$, a wino mass of $300 \hbox{~GeV}$ would correspond in the mirage scenario to $m_{\tilde{g}} \sim
570 \hbox{~GeV}$ and in the large volume scenario to $m_{\tilde{g}} \sim 900 \hbox{~GeV}$.
While the gluino mass might be difficult
to measure directly, in both models the gluino mass is correlated with the
squark masses, which are easier to
measure. Thus by measuring the
$m_{\tilde{\chi}_2^0} - m_{\tilde{\chi}_1^0}$ and $m_{\tilde{q}_L} -
m_{\tilde{\chi}_1^0}$ mass differences it will be possible to distinguish these two models.
For example, suppose $m_{\tilde{\chi}_2^0} - m_{\tilde{\chi}_1^0}$ was
measured as $75 \hbox{~GeV}$, together with $m_{\tilde{\chi}_1^0} \gtrsim 200 \hbox{~GeV}$.
Then a measurement of $m_{\tilde{q}_L} - m_{\tilde{\chi}_1^0} \sim 250
\hbox{~GeV}$ would prefer mirage mediation models to large-volume models
and a measurement $m_{\tilde{q}_L} - m_{\tilde{\chi}_1^0} \sim 550
\hbox{~GeV}$ would prefer large-volume models to those of mirage mediation.

Compared to mirage mediation, the large volume models have a less bunched spectrum. This can be understood by the nature of the
soft term universality that arises at the intermediate scale. In the large volume models, this is approximate and is broken by
the magnetic fluxes on the brane world volumes. The effect of this flux-breaking is to raise the gluino mass in relation to the wino mass:
as the gluino mass runs up, this broadens the low-energy spectrum. In mirage mediation models the gaugino masses exhibit
exact universality at the intermediate scale, and so at low energies the $m_{\tilde{g}} : m_{\tilde{W}}$ ratio remains smaller
than for the large-volume models.

It thus should be possible to use the gaugino masses to distinguish the soft terms produced by the large-volume models from those appearing in
either mSUGRA or
 mirage mediation scenarios. Distinction from mSUGRA is possible through the $m_{\tilde{W}} : m_{\tilde{B}}$ ratio but not through
the $m_{\tilde{g}}: m_{\tilde{W}}$ ratio; distinction from mirage mediation is possible through the $m_{\tilde{g}} : m_{\tilde{W}}$ ratio
but not through the $m_{\tilde{W}} : m_{\tilde{B}}$ ratio. In both cases further distinction should be possible through the spectrum
of scalar masses. However it should be remarked that the mirage mediation
scenario is also based  on IIB string compactifications and the
presence of magnetic fluxes may also affect the structure of soft
terms in that scenario.

\section{Collider Observables}
\label{observables}

In this section we investigate the collider phenomenology that follows from the spectra given in
section \ref{spectra}. We analyse this through Monte Carlo simulation
of $10 \hbox{fb}^{-1}$ of LHC data for each spectrum. For one given model we
simulate $100 \hbox{fb}^{-1}$ of LHC data
and show how well the spectrum can be successfully reconstructed.

\subsection{Collider Model and Data Generation}

To generate events we used PYTHIA version 6.400 \cite{pythia}
linked to the PGS (Pretty Good Simulation) detector simulator \cite{PGS}.
We take the {\tt SOFTSUSY2.0} spectrum and link it to PYTHIA with the SUSY Les
Houches Accord~\cite{hepph0311123}.
For each particle spectrum generated we have simulated $10{\rm fb}^{-1}$ of
data for $pp$ collisions
at 14 TeV.
We also simulated the $t \bar{t}$ and $WW/ZZ/WZ$ Standard Model backgrounds.
We did not simulate the $W/Z + \hbox{jets}$ background due to the large amounts
of CPU time required.
Data was originally generated in the LHC Olympics \cite{LHColympics}
format. This contains
the particle energies and momenta for all particles
(electrons, muons, hadronic taus, jets and photons) in an event, and b-tagging
information for jets. The data analysis was performed using ROOT \cite{ROOT}.

The
SUSY production cross sections for a given spectrum can be computed using
\verb+Prospino 2.0+ \cite{Prospino} at next-to-leading order in QCD. With a
gluino scale $m_{\tilde{g}} \sim 900$ GeV and the spectra of
figure \ref{massesM3is500} and \ref{masses40percentM3is500}, the dominant
production cross-sections are $\sigma_{\tilde{g} \tilde{g}} \sim 0.53-0.54$
pb,
$\sigma_{\tilde{q}\tilde{g}} \sim 2.9-3.0$ pb and
 $\sigma_{\tilde{q}\tilde{q}} \sim 1.4-1.5$ pb.
Squark-gluino production is therefore most significant,
followed by squark-squark and then gluino-gluino production.

Using PYTHIA and PGS, we simulated $10 {\rm fb}^{-1}$ of mock LHC data for each of the spectra that were
generated in section \ref{spectra}. The basic cuts used were the PGS Level 2 triggers,
which are summarised in the appendix.
As all of these models have a similar SUSY production scale and cross-section,
differences in the number of events passing the cuts must be attributed to the
detailed structure of the spectrum rather than to the overall scale.
To compare and contrast many different models, all with the same high-energy origin and with the same
overall scale, we first use counting observables.
These will probably not be very useful for discovery of physics beyond
the Standard Model, being too sensitive to experimental systematics and
physics unknowns such as parton densities and parton shower/matrix element
approximations. Nevertheless, we may imagine a time in the future when these
effects have been measured to some precision at the LHC after a beyond the
Standard Model discovery, and counting observables might be used for model
discrimination. Their importance here is that they provide a simple
and easily visualisable measure of the differences in
the possible phenomenology across a wide range of models.
Although not our emphasis in this study, the observables used are
mostly chosen to be those
in \cite{hepph0610038}.\footnote{The results given for the large-volume models
  in
\cite{hepph0610038} differ from those here as the soft terms used are different:
in \cite{hepth0610129} it was shown that a cancellation exists in the scalar
soft terms
that significantly reduced the scalar soft terms by a factor $\ln(M_P/m_{3/2})$ compared to
the original estimate of \cite{hepth0605141} that was used in
\cite{hepph0610038}.
The K\"ahler potential was calculated for
Calabi-Yau manifolds to first order in a volume expansion only
recently~\cite{hepth0609180}.
The cancellation occurs only for chiral fields.}
While the cuts employed here are somewhat different to those used in
ref.~\cite{hepph0610038}, a rough comparison between their results and ours
may be useful.
We also investigate the effect of increasing the
high scale parameter fluctuations as well as varying the overall scale of the soft terms.

In section \ref{reconstructionsubsec} we discuss the potential to reconstruct the spectra from kinematic observables.
We will find that this is much easier for some spectra than for others.

\subsection{Counting Observables}
\label{subsecCounting}

A counting observable is simply the number of events in a sample which all
satisfy a certain set of desired properties. Since the
expected number of
events $N$ with properties $\cal{P}$ has a statistical uncertainty
$\sqrt{N}$, the signal is taken to be observable only if it is well
above a large statistical fluctuation of the background. Therefore for
observability we require
\begin{equation}
\label{countdef}
{N_s\over{\sqrt{N_b}}} > 4, {N_s\over{N_b}} > 0.1, N_s > 5,
\end{equation}
where $N_s$ is the number of events in the signal data sample
satisfying property $\cal{P}$ and $N_b$ is the number of events in the
background sample satisfying the same property. Our estimation of $N_b$ could
easily be wrong by factors of a few due to theoretical and experimental
uncertainties, but we shall see that thankfully most observables will still
be usable.

\begin{figure}
\begin{center}
\includegraphics[width=12cm]{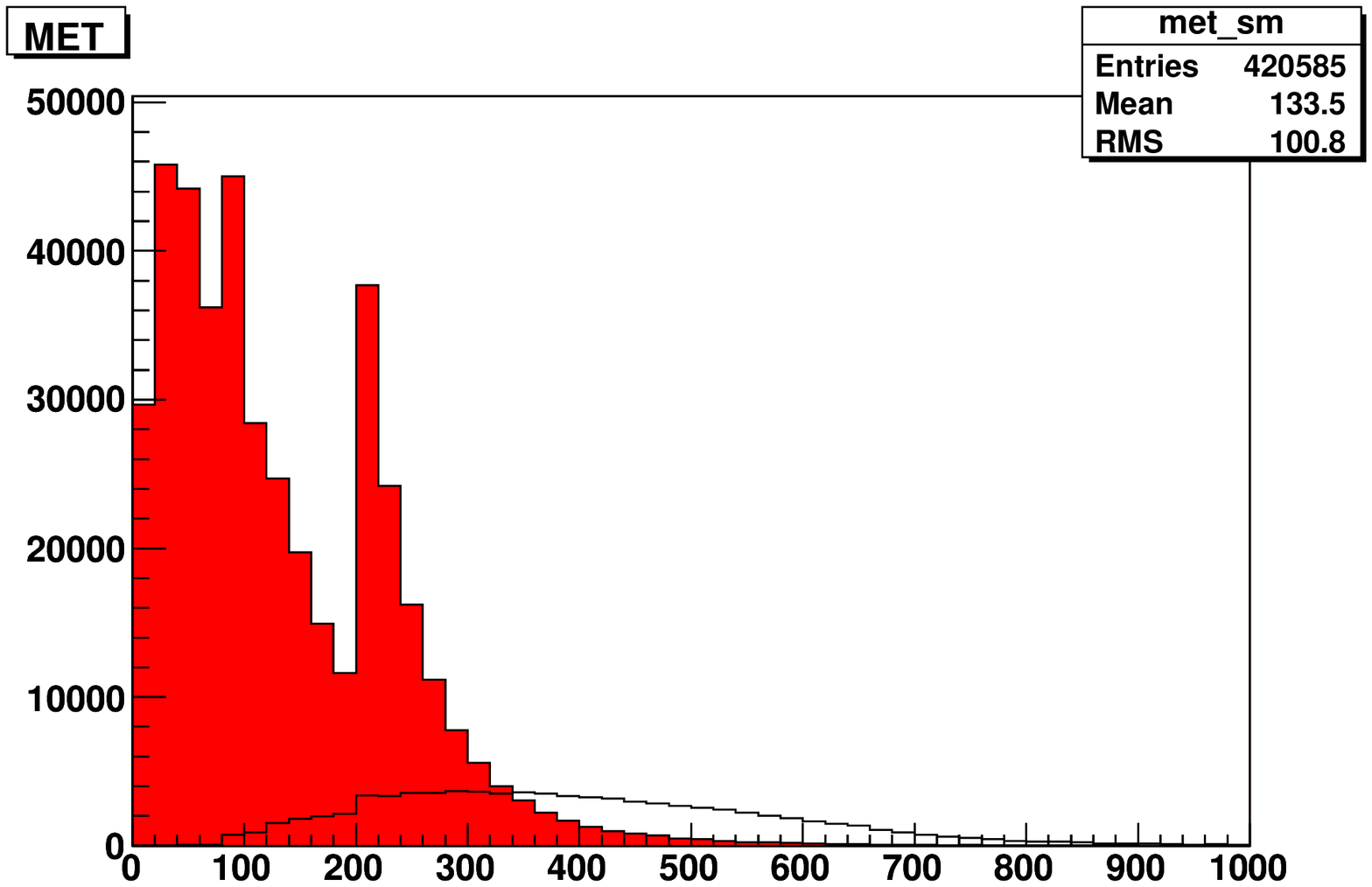}
\caption{$\MET$ plot for background and a sample signal spectrum. The background is
 shown in red. The spike in the background $\MET$ at 200 GeV is an artifact of the triggers used.
 \label{metplot}}
\end{center}
\end{figure}

The basic cuts for the counting observables are the L2 triggers
used in the LHC Olympics version of PGS.  On top of that we impose
additional cuts:
\begin{enumerate}
\item For all jets entering the counting, we require jet $P_T > 100$ GeV.
\item For all isolated $e,\mu$, we require $P_T > 10$ GeV.
\item For all $\tau$s require $P_T > 100$ GeV.
\item Missing transverse momentum $\MET > 300$ GeV.
\end{enumerate}
The choice of $\MET$ cut can be motivated by the plot in figure
\ref{metplot}, produced for one of the randomly generated spectra. It
can be seen
that for $\MET>340$ GeV, the signal dominates over the
background. This value does not depend significantly on the
choice of the random spectrum.

The simplest signatures to consider are inclusive counts of opposite
sign di-lepton and tri-lepton events (including taus). These are also
among the most promising for SUSY searches since the detector
tagging efficiency for leptons (electrons and muons in particular)
is quite high. Furthermore, there is relatively little Standard Model background for
multi-lepton multi-jet high-$\MET$ events.

The presence of many opposite sign di-leptons is also an indicator of
whether the decay chain
$\tilde{\chi}_2^0\to \tilde{l}^{\pm} l^{\mp} \to l^{+} l^{-} \tilde{\chi}^0_1$
occurs frequently in the samples. This decay chain may be
used~\cite{ATLASTDR} for constructing
an $l^+ l^-$ endpoint which can subsequently be used to constrain sparticle
masses.
The results for spectra with $m_{\tilde g}$ fixed at $\sim 900$ GeV are
shown in Figure \ref{obs23M3is500}.
\FIGURE[r]{
\includegraphics[width=7cm]{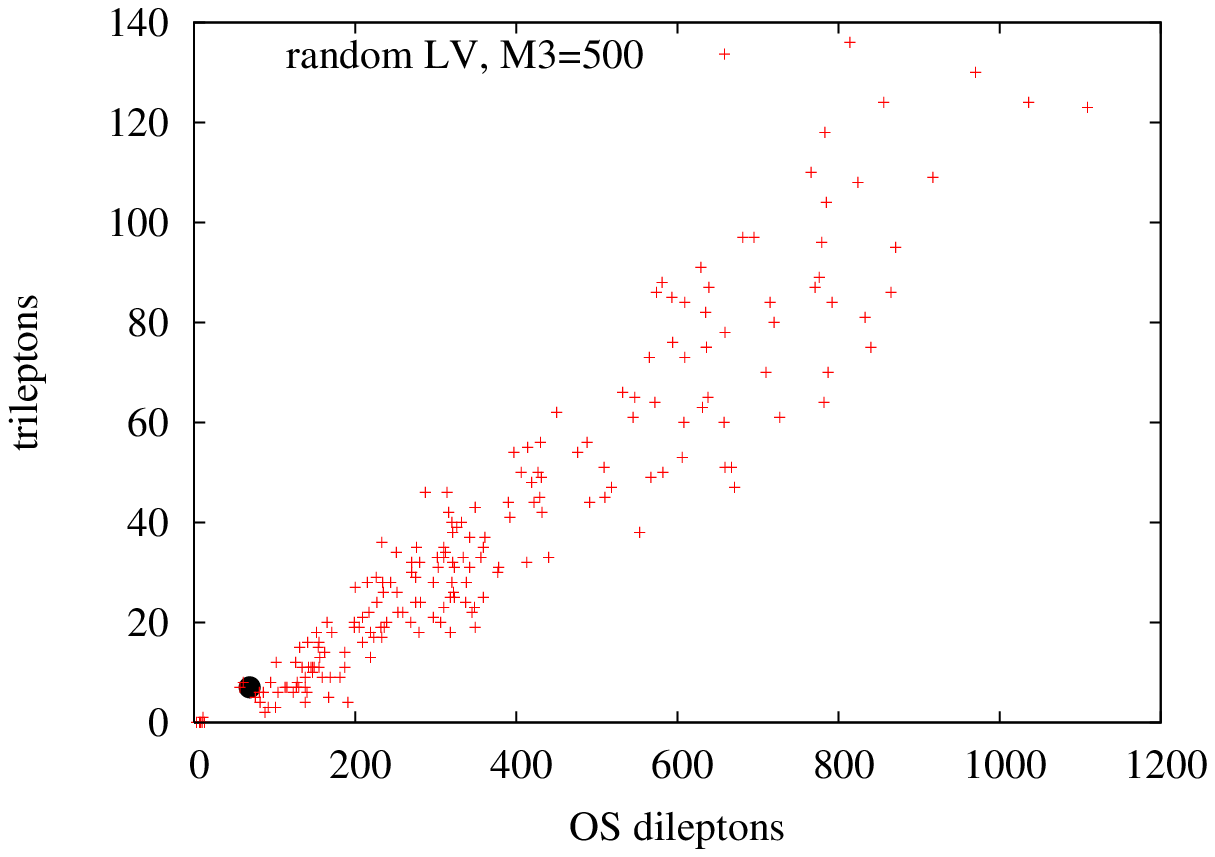}
\caption{Numbers of tri-leptons/OS di-leptons. The black dot corresponds to the observable
limit according to (\ref{countdef}). \label{obs23M3is500}}
}
Even though the overall mass scale of the spectrum is fixed and the SUSY cross-section is essentially unaltered
between models, there is still a
large range for the number of observed di-lepton and tri-lepton events.
The dominant background comes from $t\bar{t}$ events, and
in a large number of cases the number of such SUSY events lies above the
acceptable number of background events.

This variation can be understood in terms of the different possible spectra,
The number of di-lepton (and hence tri-lepton) events depends crucially
on the details of the $\tilde{\chi}_2^0\to \tilde{l}^{\pm} l^{\mp} \to
l^{\pm} l^{\mp} \tilde{\chi}_1^0$ decay chain. In most cases the first 2
generations of left handed sleptons are heavier than
$\tilde{\chi}_2^0$.  However the right handed sleptons may be
lighter or heavier than $\tilde{\chi}_2^0$. If they are all heavier, then the
3-body decays of the $\tilde{\chi}_2^0$ will dominate. The $\tilde{\chi}_2^0 \to \tilde{\chi}_1^0$ decay will
then often proceed through an off-shell slepton $\tilde{l}^*$, giving many
di-lepton events. However, if $\tilde{e}_R, \tilde{\mu}_R$ are
heavier than $\tilde{\chi}_2^0$ but $\tilde{\tau}_1$ is lighter - which can
occur as, for larger $\tan \beta$, $m_{\tilde{\tau}_1}$ is driven down by the
RGEs and tends to be light - the
$\tilde{\chi}_2^0$ will predominantly decay
through the chain $\tilde{\chi}_2^0 \to \tilde{\tau}^\mp \tau^{\pm} \to \tilde{\chi}_1^0 \tau^{\mp} \tau^{\pm}$.
Since the tau tagging efficiency is relatively low, and $m_{\tilde{\tau}} - m_{\tilde{\chi}_1^0}$ is relatively
small (meaning that the taus tend to be soft), few taus will be picked up by the
detector.
It may also happen that all of $\tilde{e}_R, \tilde{\mu}_R,
\tilde{\tau}_R, \tilde{\nu}$ are lighter than $\tilde{\chi}_2^0$, with the
dominant branching ratio of $\tilde{\chi}_2^0$ again into $\tilde{\tau}$s and $\tilde{\nu}$s.
In this case we again observe only a few di-lepton or tri-lepton events.

In ref.~\cite{hepph0610038}, the number of clean di- and tri- leptons was used
as an observable, meaning no jet activity in the detector. From the point of
view of our simulation, this corresponds to direct weak gaugino and/or slepton
production.
In 10 fb$^{-1}$,
none of our model samples produced a single ``clean'' event,  consistent with
the masses of
the sleptons/weak gauginos and the fact that this is a weakly interacting
production channel.
The usefulness of this observable is questionable, since
there will always be some jet pollution
 in the detectors due to the QCD
background. Thus some cut on hadronic activity must be given experimentally
in order to define a jet veto, and the predicted backgrounds can be notoriously
unreliable.

Another observable considered in \cite{hepph0610038} that we will not use
here concerns the number of events with no leptons, 1 or 2 $b$-jets, and
at least six hard jets. The difficulty in using this observable comes in the estimation of the
background. The processes in PYTHIA are $ 2 \to 2$ rather
than $2 \to \textrm{many}$, and so the `hard jet' background arising
from PYTHIA comes from the parton shower rather than from direct hard jet
production. This may underestimate the background by orders of magnitude.
 A correct estimate of this background would require the inclusion of $2 \to 4,5,6$ processes in
 the Monte Carlo (e.g. with ALPGEN \cite{alpgen}).
\FIGURE[r]{
\includegraphics[width=7.5cm]{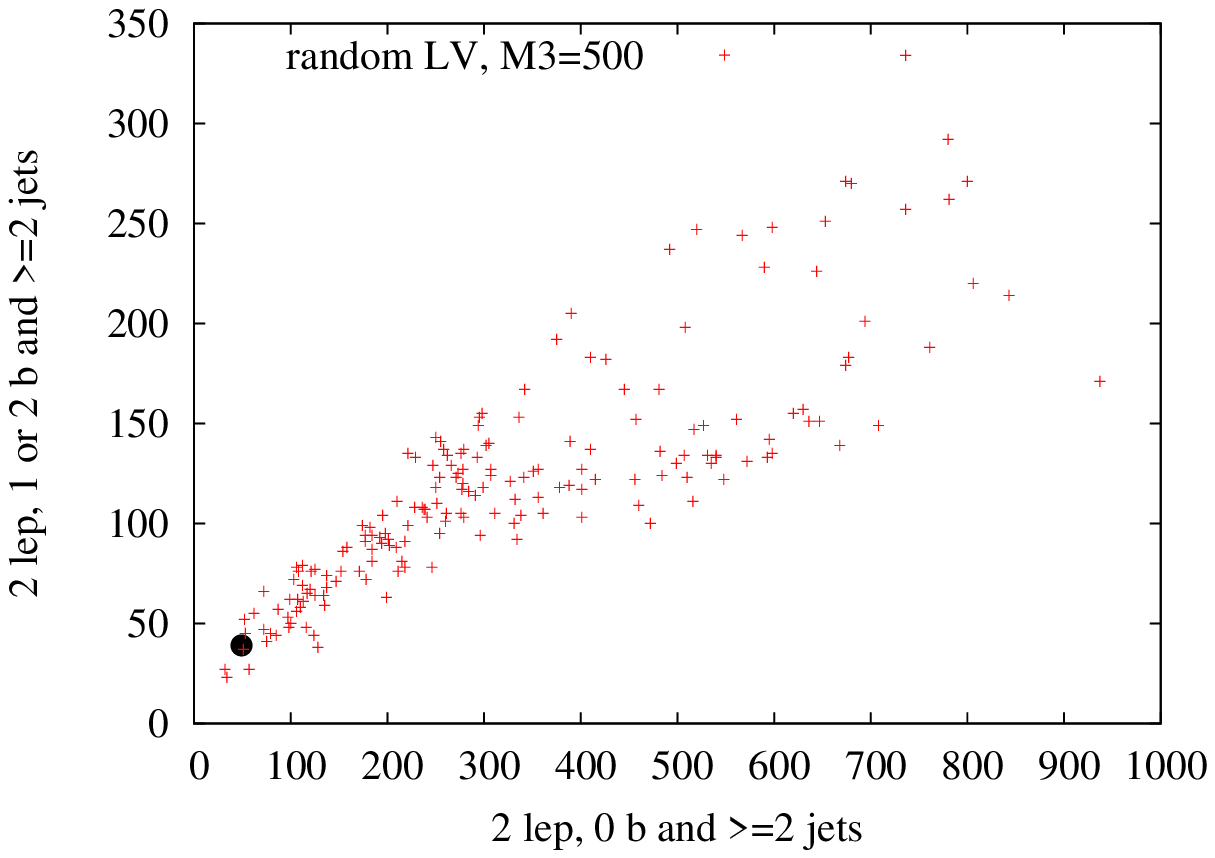}
\caption{Number of events with 2 leptons, 1 or 2 $b$-jets and 2 jets/Number of events with
2 leptons, 0 $b$-jets and 2 jets. The black dot in both case represents the observable
limit according to (\ref{countdef}). \label{obs45M3is500}}
}

In figure \ref{obs45M3is500}, we plot the number of di-lepton multi-jet
events with 1 or 2 $b$-jets against the number of di-lepton multi-jet events with 0 $b$-jets.
There is a clear positive correlation in Fig.~\ref{obs45M3is500}, indicating
that
the number of events with two leptons and $b$-jets is dependent
more on the number of events with leptons passing the cuts than on the number
of $b$-jets. Nonetheless the number of events with $b$-jets can vary by a
factor of 2 between different spectra.
This is reasonable, because there
  are numerous competing decays where the branching ratios are dependent on
  various independent input parameters. We give one example. If kinematically
  allowed, as here, the gluino has a significant branching ratio to $t {\tilde t}_1$,
  which generates events with multi-jets and $b$s.  The branching ratios of
  ${\tilde t}_1 \rightarrow t \tilde{\chi}_{1,2}^0$ and $b \tilde{\chi}_1^+$
  depend on $\tan \beta$ and
  $A_t$, since they affect the mixing of the stops. In figure \ref{obs45M3is500} we also
require two leptons. However the number of leptons observed depends on the mass differences between the
light neutralinos and the sleptons.
The point is that these parameters mentioned above are varied independently in our
models, which explains why spectra with similar numbers of leptons can have different numbers of
$b$-jets.
Other decay chains, for example $\tilde{g} \to \bar{b}\tilde{b}$, can be
analysed in a similar fashion.
\begin{figure}
\twographs{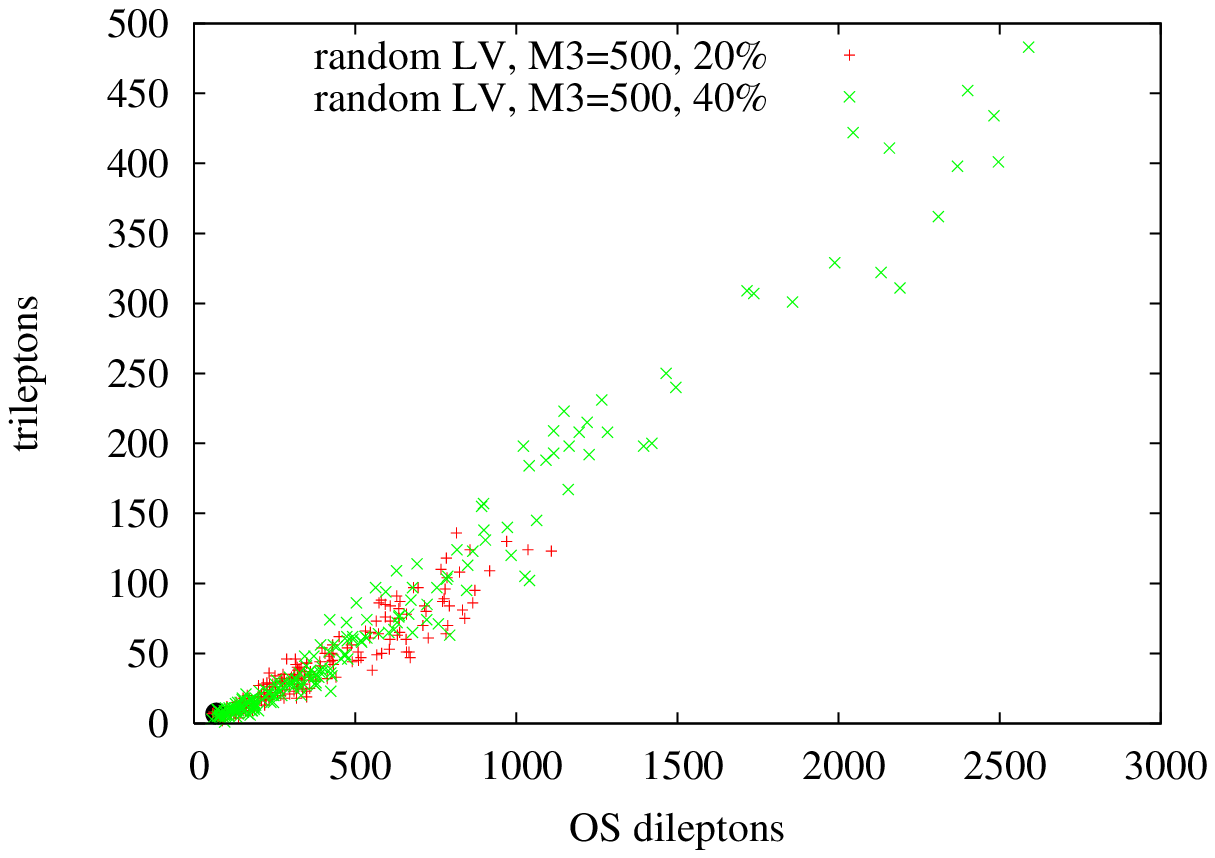}{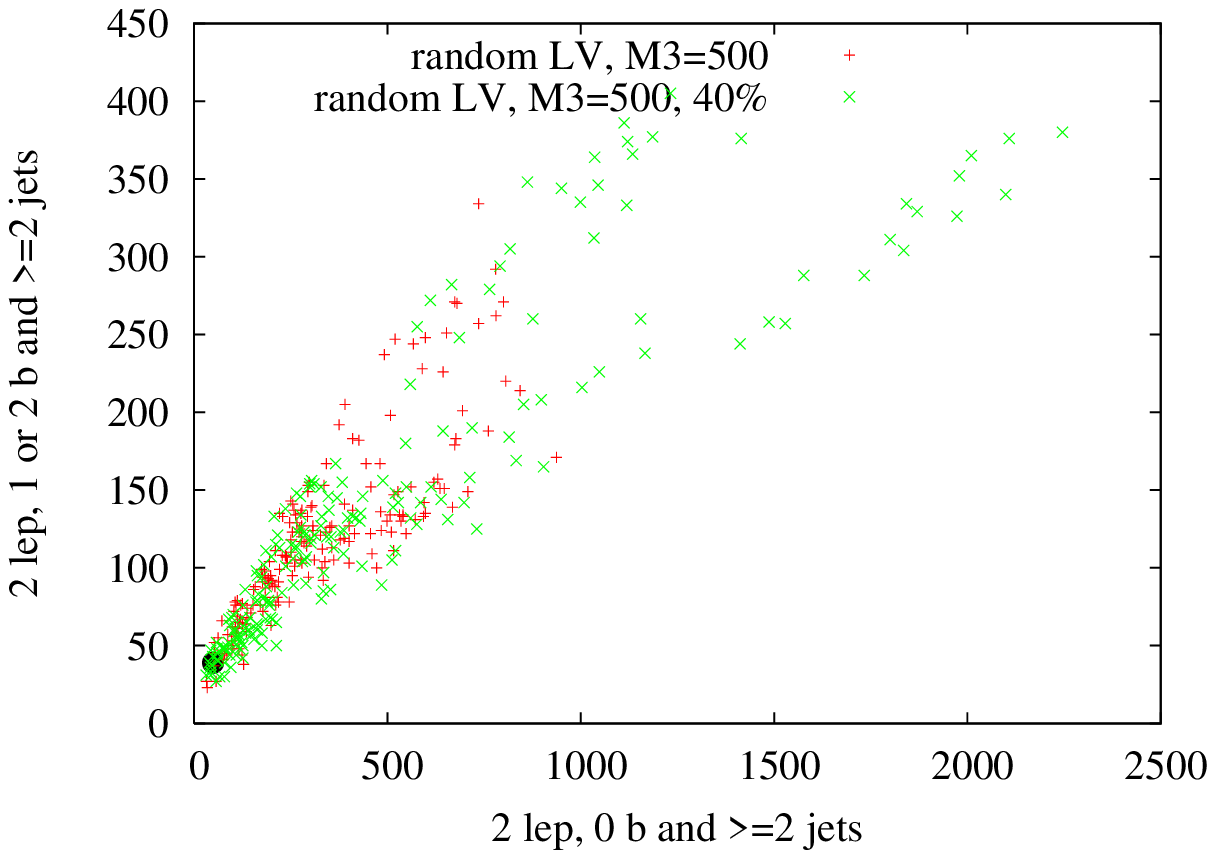}
\caption{Comparison of results with $20\%$ and $40\%$ variations: di- and tri-leptons counts and events with
$b$-jets.\label{comp1}}
\end{figure}

The number of $b$-jets may potentially be used as a coarse discriminator between different high-scale constructions.
Suppose the scalars are much heavier than the gluino, with a large Higgsino component to the LSP.
In this situation, the gluino will primarily decay through a 3 body channel to $b \bar{b} \tilde{\chi}_{1,2}^0$
- as well as other channels involving tops if kinematically allowed - due to the large higgsino coupling to stops and sbottoms.
These decays will all result in a large number of $b$-jet events. An example of such a construction is the focus point region
of mSUGRA. Since in our construction the light neutralinos are always gaugino dominated, the number of $b$-jet events is expected to be smaller.
We have verified that a focus point spectrum does indeed give many more $b$-jet events than those arising from
the large-volume models. Thus, once $b$-jet tagging is understood,
the number of $b$-jets can potentially be used to distinguish different
high-scale constructions with a
similar overall production scale.

\FIGURE[r]{
\includegraphics[width=7.5cm]{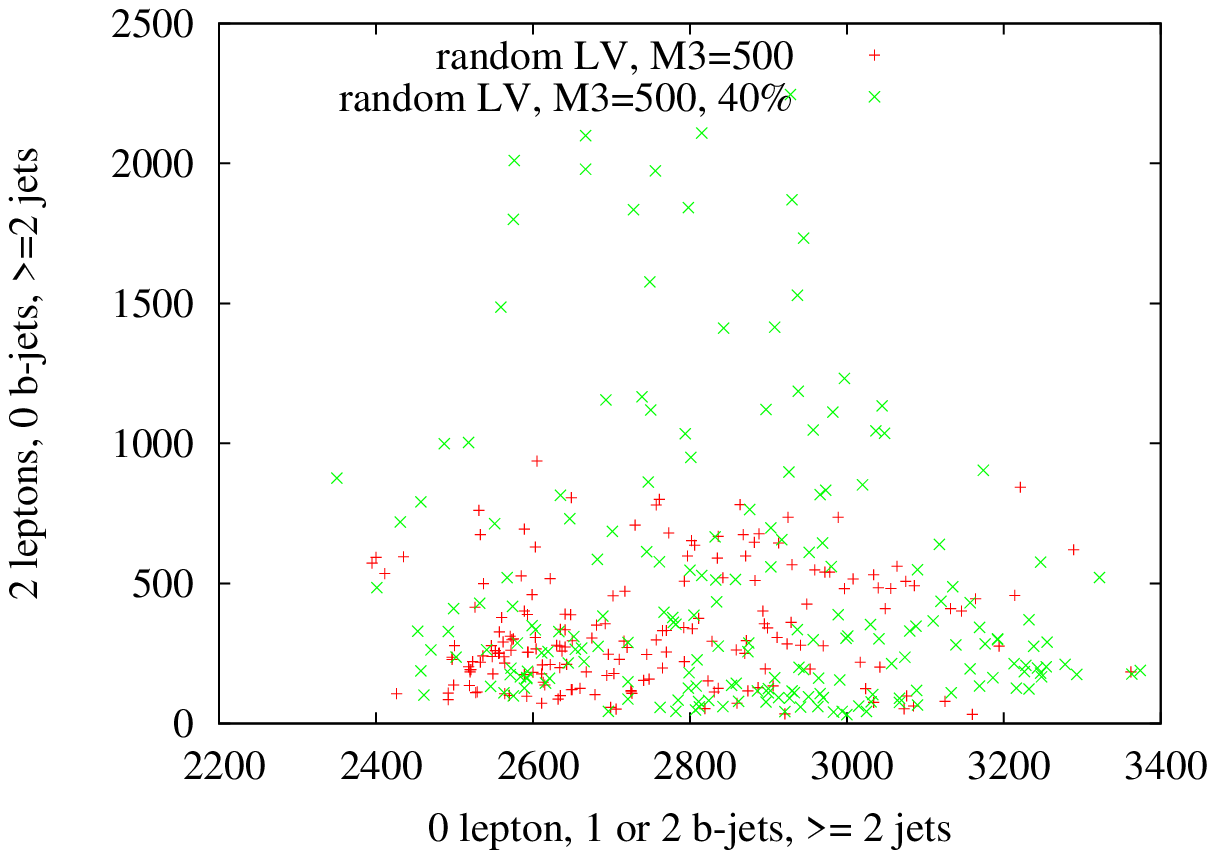}
\caption{Number of events with 0 leptons, 1 or 2 $b$-jets and 2 jets vs.\
  number of   events with 2 leptons, 0 $b$-jets and 2 jets.
\label{comp2}}}
Since the $20\%$ random variation we chose for all parameters is an
arbitrary choice, it is
  useful to compare the results with a
sample of spectra with $40\%$ variations. The results (again with
$M_3=500$ GeV at the high scale) are shown in figures \ref{comp1}-\ref{comp2}.
In figure \ref{comp1} we see the number of di- and tri-lepton events, as well as the number of di-lepton events
with or without $b$-jets. The general structure of the observables is similar to that seen above
with the $20 \%$ variations.
However the $40\%$ variation spectra, although with the same overall mass scale
(as defined by the gluino mass) can have significantly more di-lepton events
than the $20\%$ variation spectra. This is due to the fact that in the
$40\%$ case, the LSP is more likely to acquire a significant wino component, in which case the large
left-handed couplings result in a lot of lepton production
Also it may happen that all the right handed sleptons and left handed
sleptons are lighter than $\tilde{\chi}_2^0$, and that the mass differences are
all large, which again results in many observed di-lepton events.

An interesting counting observable to consider is the
number of events with one or two $b$-jets, requiring 2 hard jets in the
event. In figure \ref{comp2} we compare the number of di-lepton events without $b$-jets against the number of
0 lepton events with $b$-jets.
This observable is shown on the abscissa of figure \ref{comp2}, and it is
consistent throughout the entire sample of spectra. The observability limit
defined by Eq.~(\ref{countdef}) is at (209,49) and is omitted from the figure.

\subsubsection{Varying the Sparticle Mass Scale}

\begin{figure}
\twographs{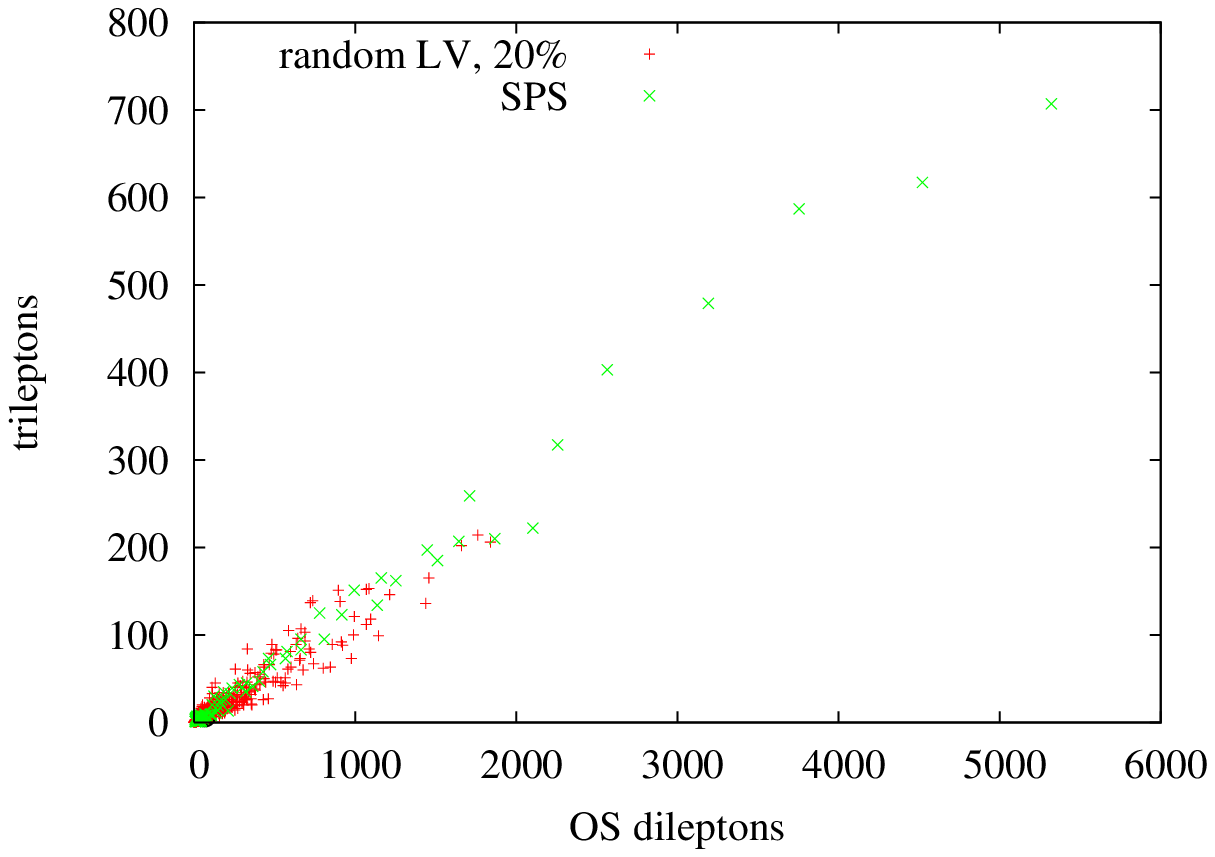}{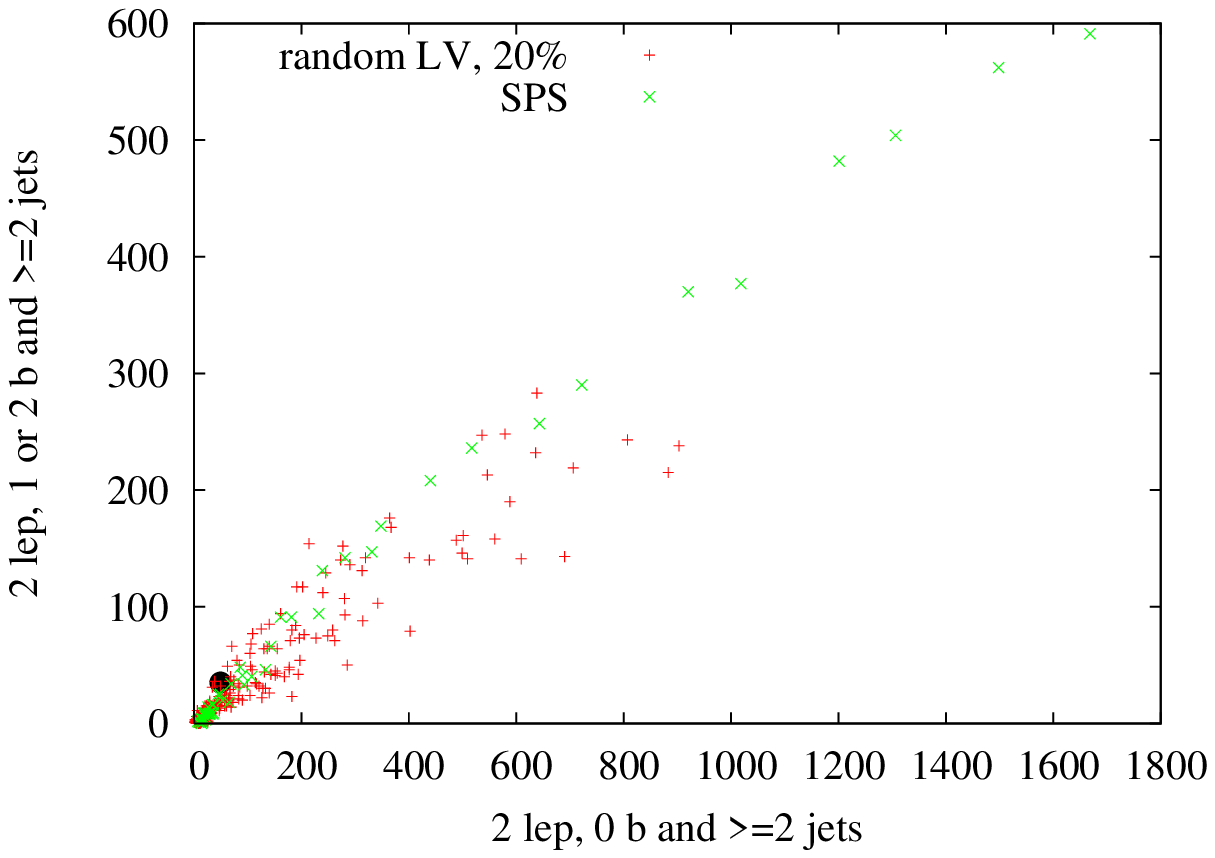}
\caption{Comparison of event numbers with $20\%$ variations and varying mass
scale, and results for the SPS1a slope:
(a) numbers of tri-leptons/OS di-leptons, (b) number of events
with 2 leptons, 1 or 2 $b$-jets and 2 jets/number of events with
2 leptons, 0 $b$-jets and 2 jets.
\label{SPS1}}
\end{figure}
We next investigate the effect of varying the overall mass scale which
has so far been set by fixing the gluino mass. An arbitrary value for
$M_3$ is selected and $20\%$ variations on the parameter
$F^s/(\tau_s)$ allowed. The results are shown in figures
\ref{SPS1}-\ref{SPS3}, in conjunction with the results for the SPS1a
slope given by $m_0 = 0.4 m_{1/2}, A_0 = - m_0, \mbox{sgn}\mu=+1, \tan\beta=10$.
In the plots, 50 SPS1a points from $M_{12}=250$ GeV to $M_{12}=887$ GeV in
steps of 13 GeV are shown.
\FIGURE[r]{
\includegraphics[width=7.5cm]{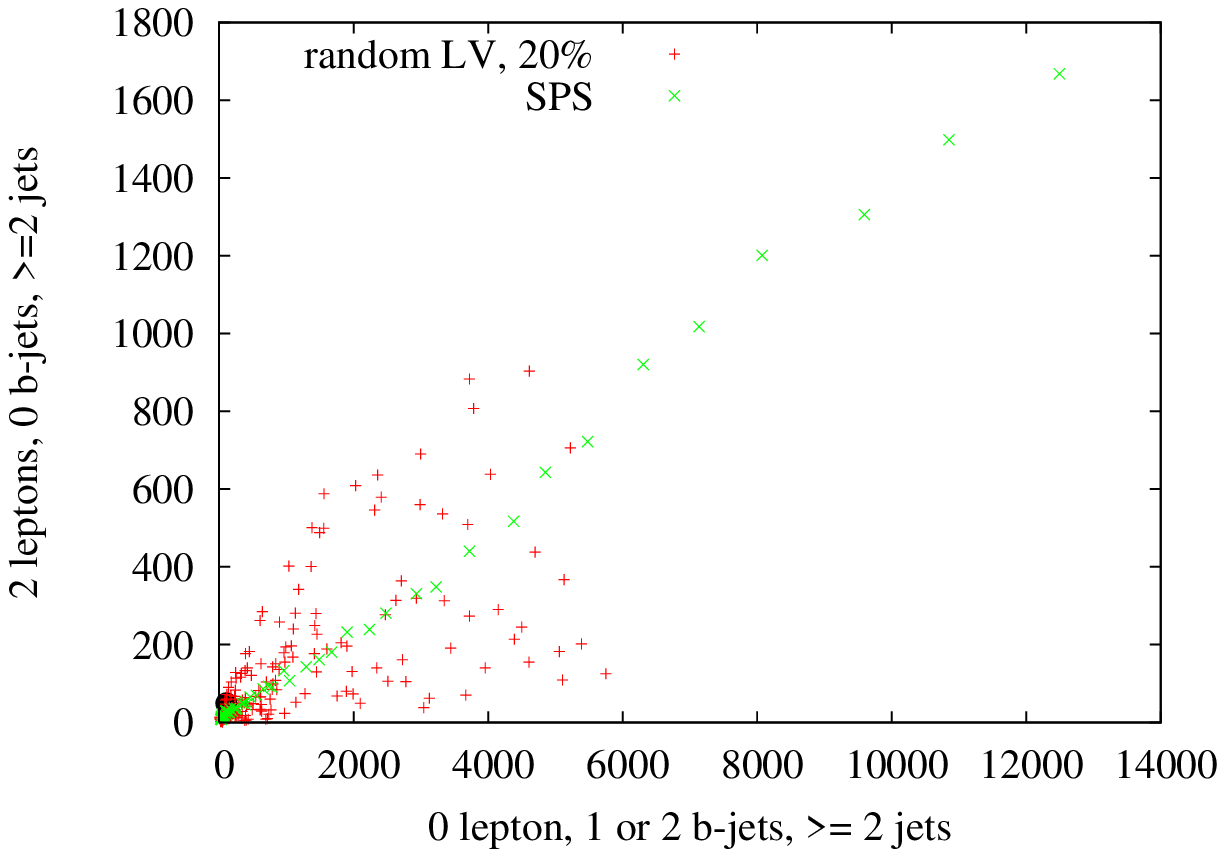}
\caption{Comparison of event numbers with $20\%$ variations and a varying mass
scale. The results for the SPS1a slope are included for comparison.
\label{SPS3}}}
The main conclusion from the study of counting observables is that,
 even with the spectra restricted to the form
of figures \ref{massesM3is500} and \ref{masses40percentM3is500}, the total number of observed SUSY events can vary
 widely.
Even fixing the overall scale of
the spectrum, the large-volume models still lead to a widely varying
 number of triggered
events: models at the same
scale and coming from the same high-scale theory can give quite
 different results
  as small changes in high scale parameters can lead to
 significant changes in observables.
We do not make an explicit comparison of our results with the models
 discussed in \cite{hepph0610038} for this reason. If there is so much variation
 within our models, there will be a larger variation with other models
 that would make them difficult to differentiate. We have also used different
 cuts, which makes a quantitative comparison impossible. Nonetheless, a rough look at
 the similar plots in both works does not indicate an
 easy way to
 separate these models from those in \cite{hepph0610038}.

\subsection{Potential for Reconstruction}
\label{reconstructionsubsec}

We now discuss the potential for reconstructing the spectra of figure \ref{massesM3is500}. This will illustrate the above
point: if there are very few di-lepton events, and no kinematic endpoints, then
direct reconstruction would be very difficult if not impossible. The structure of the analysis given here
follows standard accounts of reconstruction such as in ref.~\cite{ATLASTDR,hinchliffe},
for example.

\begin{figure}
\begin{center}
\twographs{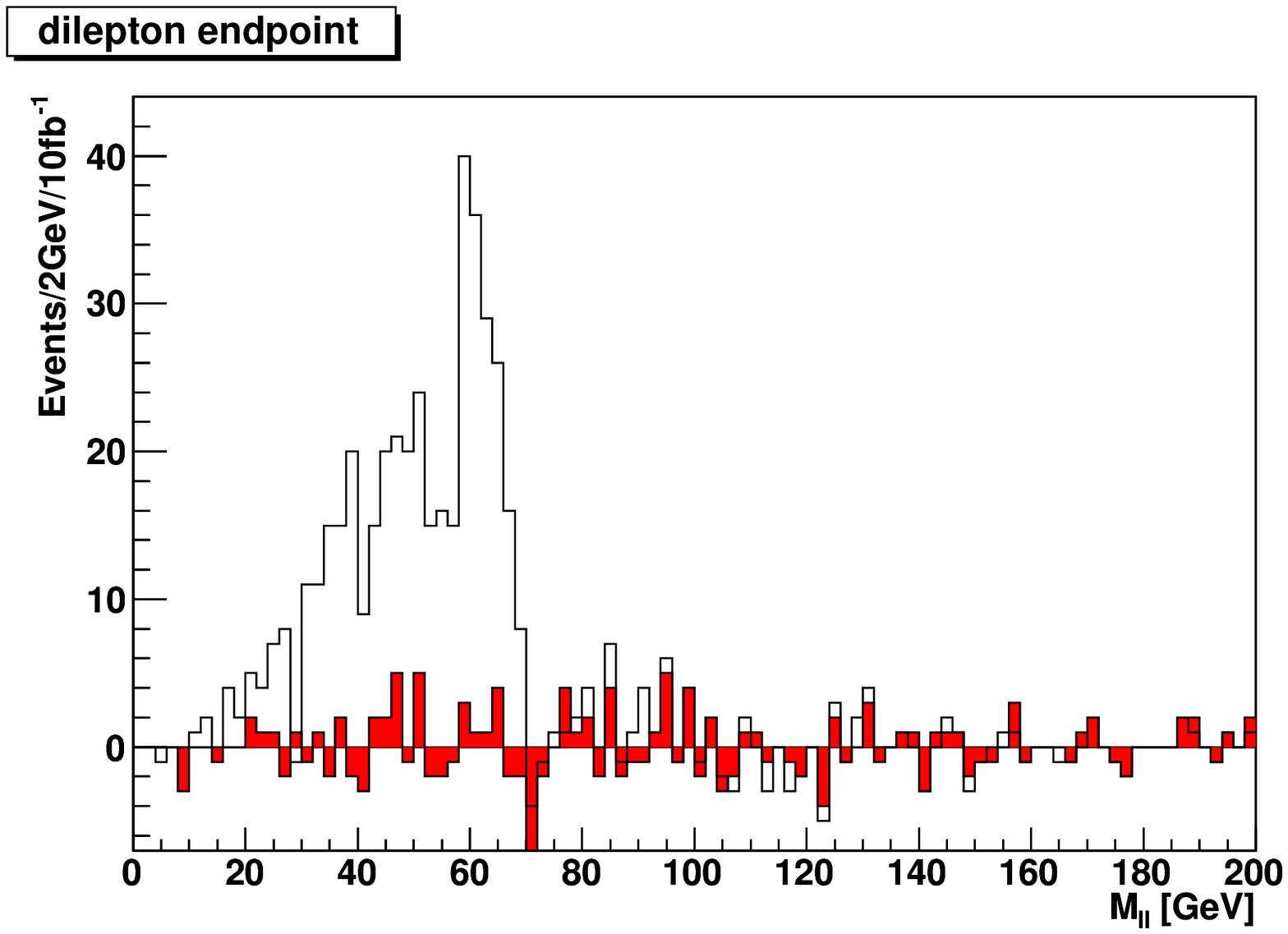}{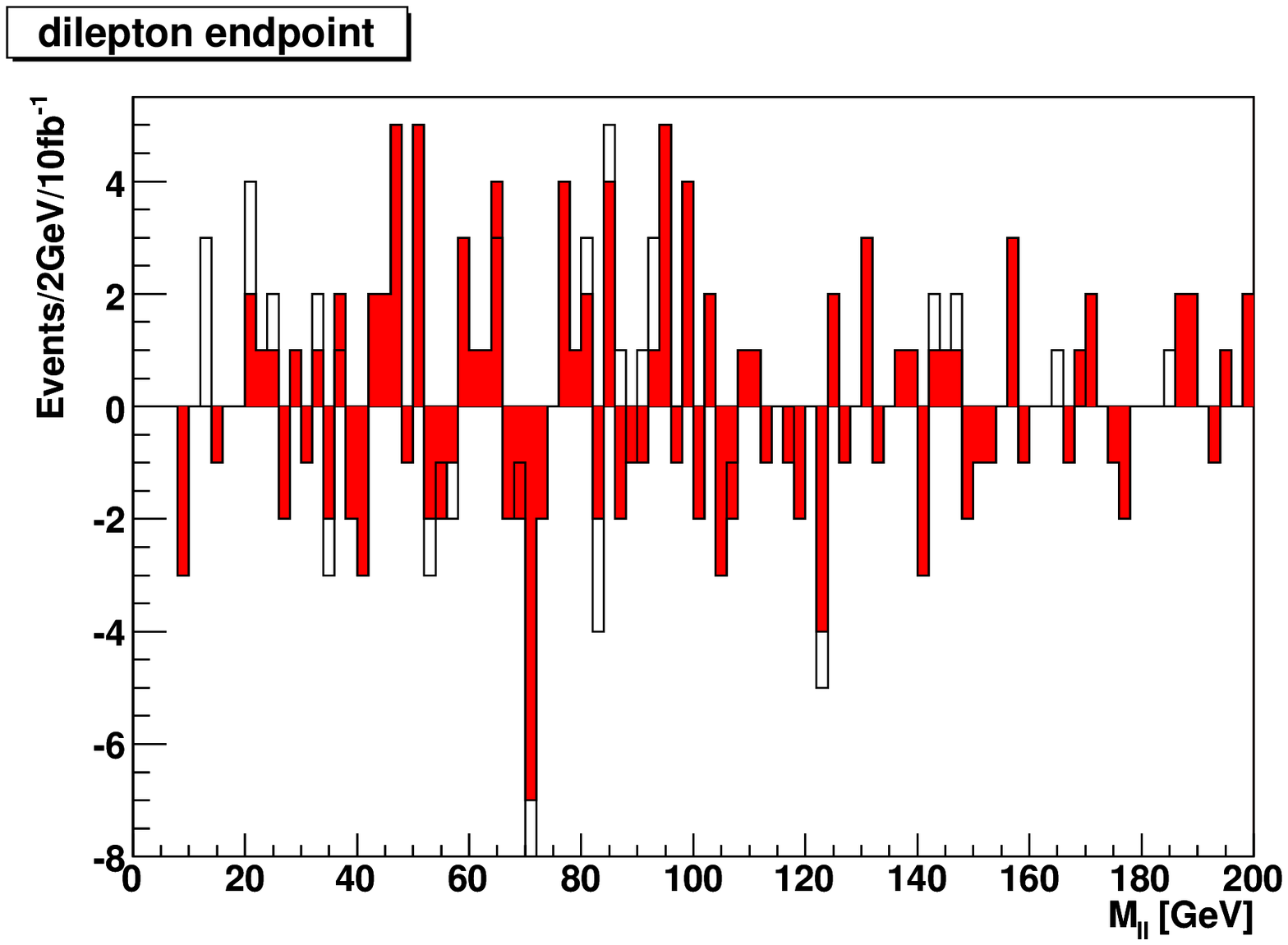}
\caption{Opposite sign, same flavour di-lepton invariant mass
for (a)  a spectrum with many di-lepton events
and $m_{\tilde{\chi}_2}\gg m_{\tilde{\chi}_1}$ and (b) a spectrum
with few di-lepton events and
$m_{\tilde{\chi}_2^0}-m_{\tilde{\chi}_1^0}\approx 21$ GeV. In both cases
we require 4 hard jets and $\MET \ge 200$ GeV. Standard Model background is
shown in red.}
\label{leptonslledge}
\end{center}
\end{figure}

The ability to reconstruct supersymmetric particle masses depends significantly on the spectrum and
on the decay chains and their branching ratios.
Since jet observables usually suffer from large combinatorial
backgrounds, the cleanest measurements are those involving
only leptons. Therefore the first step in reconstruction of a
supersymmetric spectrum is a
measurement of a di-lepton edge from the $\tilde{\chi}_2^0 \to \tilde{\chi}_1^0 l^{\pm}
l^{\mp}$ chain. As explained in section \ref{subsecCounting},
whether we observe few or many di-lepton events
from this decay chain is determined by
the mass difference between $\tilde{\chi}_1^0$ and $\tilde{\chi}_2^0$ and the slepton branching ratios.
Figure \ref{leptonslledge} shows
the plot of the di-lepton invariant mass
$M_{ll} \equiv (p_{l_1} + p_{l_2})^\mu (p_{l_1} + p_{l_2})_\mu$ for 10${\rm
  fb}^{-1}$ of data for two spectra, one
with many opposite sign, same flavour (OSSF) di-lepton
events in the signal, and one with very few.
There is no evidence for an edge in the di-lepton invariant mass in
Fig.~\ref{leptonslledge}b. If there are few di-lepton
events, the spectrum is much harder to reconstruct since one has to
resort to multi-jet observables.
We therefore restrict to considering a spectrum which generates
 many OSSF di-lepton events, such that the di-lepton
edge can be easily reconstructed. For this spectrum we simulated
$100{\rm fb}^{-1}$ data and the plots given are based on this simulation.
This represents one year of LHC running at design luminosity. The spectrum we
 attempt to reconstruct is as shown in Table~\ref{tab:masses}.

\TABULAR[t]{|cccccccccc|}{\hline
$m_{\tilde{\chi}^\pm_1}$ & $m_{\tilde{\chi}^\pm_2}$ &
$m_{h^0}$ & $m_{H^0}$ &
$m_{A^0}$ & $m_{H^\pm}$ &
& &
 &
\\
303 & 480 & 114 & 532 & 532 & 538 & & & &  \\ \hline
  $m_{\tilde{g}}$ & $m_{\tilde{d}_L}$ &
  $m_{\tilde{u}_L}$ & $m_{\tilde{u}_R}$  &
  $m_{{\tilde t}_1}$ & $m_{{\tilde t}_2}$ &
  $m_{{\tilde b}_1}$ & $m_{{\tilde b}_2}$ &
  $m_{\tilde{d}_R}$  &
\\
909 & 800 & 792 & 779 & 583 & 790 & 725 & 782 & 787 & \\
\hline
$m_{\tilde{e}_L}$ &  $m_{\tilde{\tau}_1}$ &
$m_{\tilde{\nu}_{e}}$ &  $m_{\tilde{\nu}_\tau}$ &
$m_{\tilde{e}_R}$ & $m_{\tilde{\tau}_2}$ &
$m_{\tilde{\chi}^0_1}$ & $m_{\tilde{\chi}^0_2}$ &
$m_{\tilde{\chi}^0_3}$ & $m_{\tilde{\chi}^0_4}$\\
348 & 261 & 338 & 336 & 270 & 349 & 233 & 303 & 460 & 483 \\
\hline
}{Mass spectrum of model picked for
  reconstruction. All masses are listed in GeV. The first two families are
  mass degenerate. \label{tab:masses}}

We start by selecting events that pass a set of cuts that we name {\em selection A}:
\begin{enumerate}
\item $\MET > 300 \hbox{~GeV}$
\item Two opposite sign electrons or muons
with $P_T>10$ GeV.
\item At least four jets with $P_{T_{1(2)(3)(4)}} > 100(50)(50)(50) \hbox{~GeV}$.
\item $\MET > 0.2 M_{eff}$, where $M_{eff} \equiv P_{T_1}+P_{T_2}+P_{T_3}+P_{T_4}+\MET$.
\end{enumerate}
We then plot the histogram of
the di-lepton invariant mass, with a flavour subtraction of the $e^+ \mu^- + e^-
\mu^+$ result to cancel processes with leptons arising from two independent decays.

\begin{figure}
\begin{center}
\includegraphics[width=12cm]{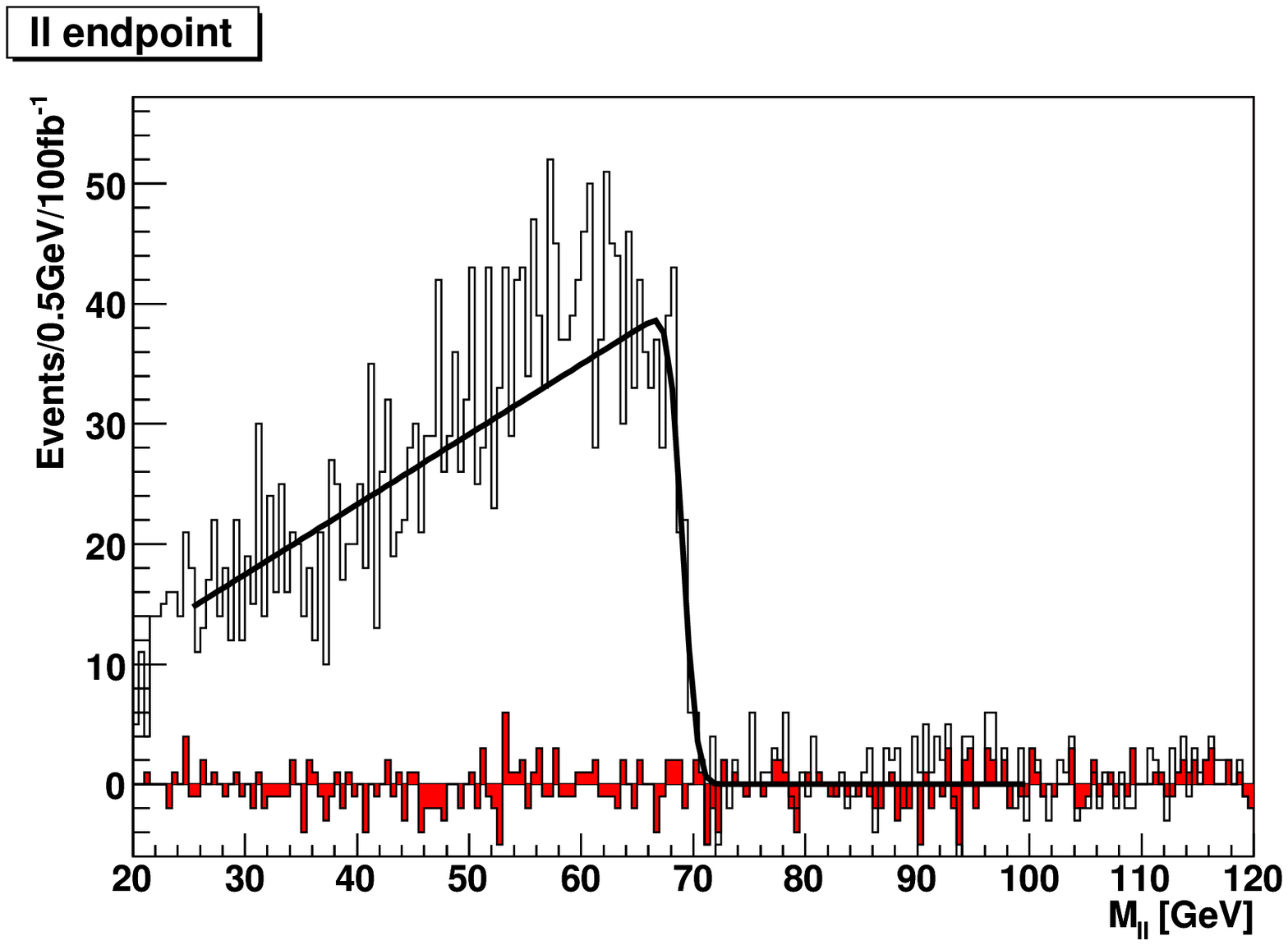}
\caption{$M_{ll}$ after cuts selection A. We expect an edge at $\sim 69$ GeV
from the spectrum. The Standard Model background is shown in red.}
\label{100fbleptonslledge}
\end{center}
\end{figure}

This plot is shown in figure \ref{100fbleptonslledge}.
It is well-known \cite{ATLASTDR}
that the $\tilde{\chi}_2^0 \to \tilde{l}^{\pm} l^{\mp} \to \tilde{\chi}_1^0
l^{\pm} l^{\mp}$ decay chain admits an endpoint at
\begin{equation}
M_{ll}^{max} = \sqrt{ { { (m_{\tilde{\chi}_2^0}^2 - m_{\tilde{l}_R}^2)
(m_{\tilde{l}_R}^2 - m_{\tilde{\chi}_1^0}^2)} \over {m_{\tilde{l}_R}^2}
}} \label{eqll}
\end{equation}
The location of this endpoint
can be found by fitting figure \ref{100fbleptonslledge} with a
triangular edge, smeared with a Gaussian (the width
of which is also fitted) to simulate resolution effects of the
experiment.
MINUIT \cite{MINUIT} and MINOS are used for this purpose, and
to estimate the error on the measurement. We obtain $M_{ll}^{max} =
69.20 \pm 0.15$ GeV.

\begin{figure}
\begin{center}
\includegraphics[width=12cm]{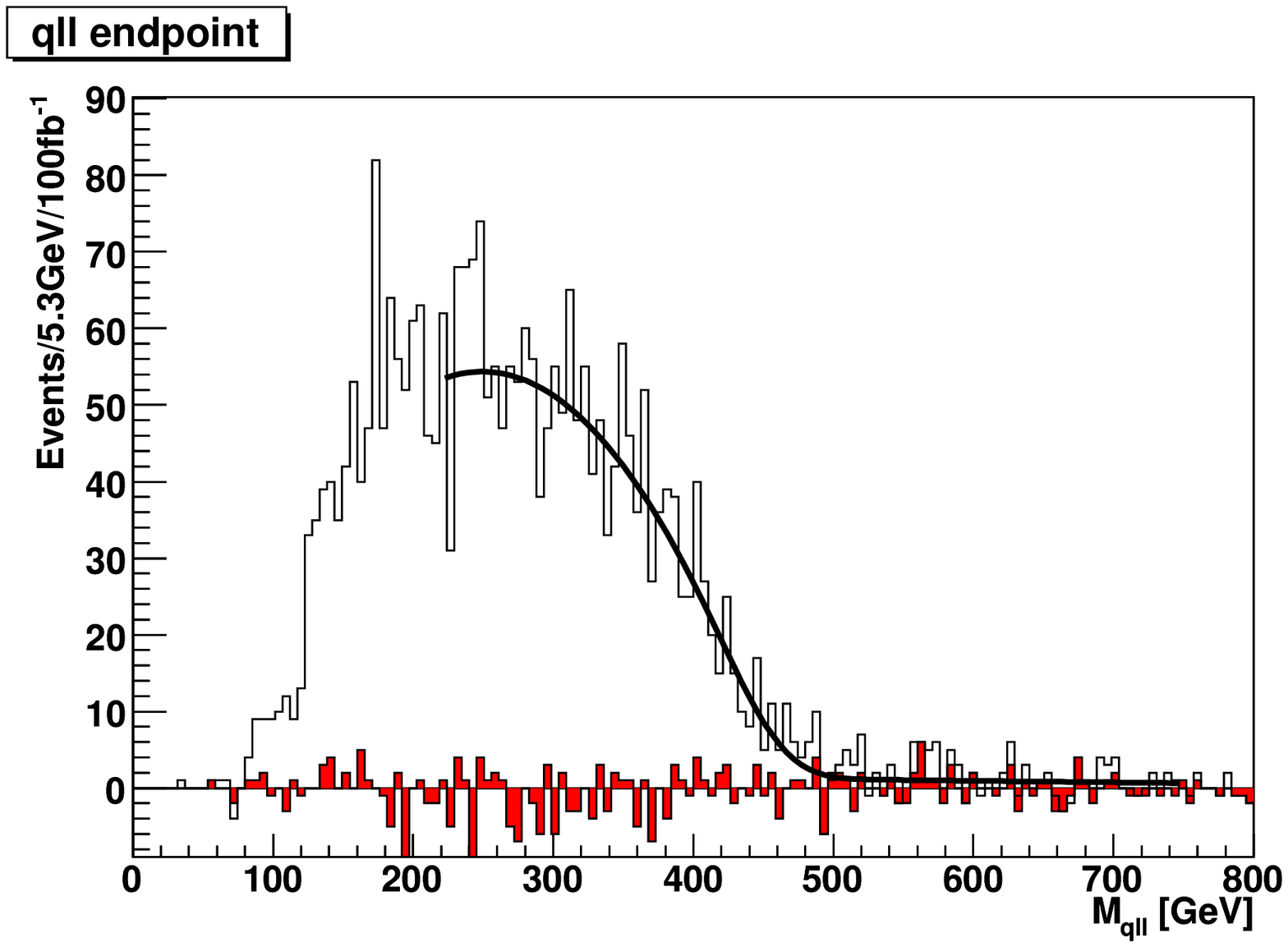}
\caption{
Reconstructing the spectrum with a clear di-lepton edge:
$M_{qll}$ edge after cuts selection B. An edge is expected at 467 GeV.
Standard Model background is shown in red. \label {qllll}}
\label{qlledge}
\end{center}
\end{figure}

\begin{figure}
\begin{center}
\includegraphics[width=12cm]{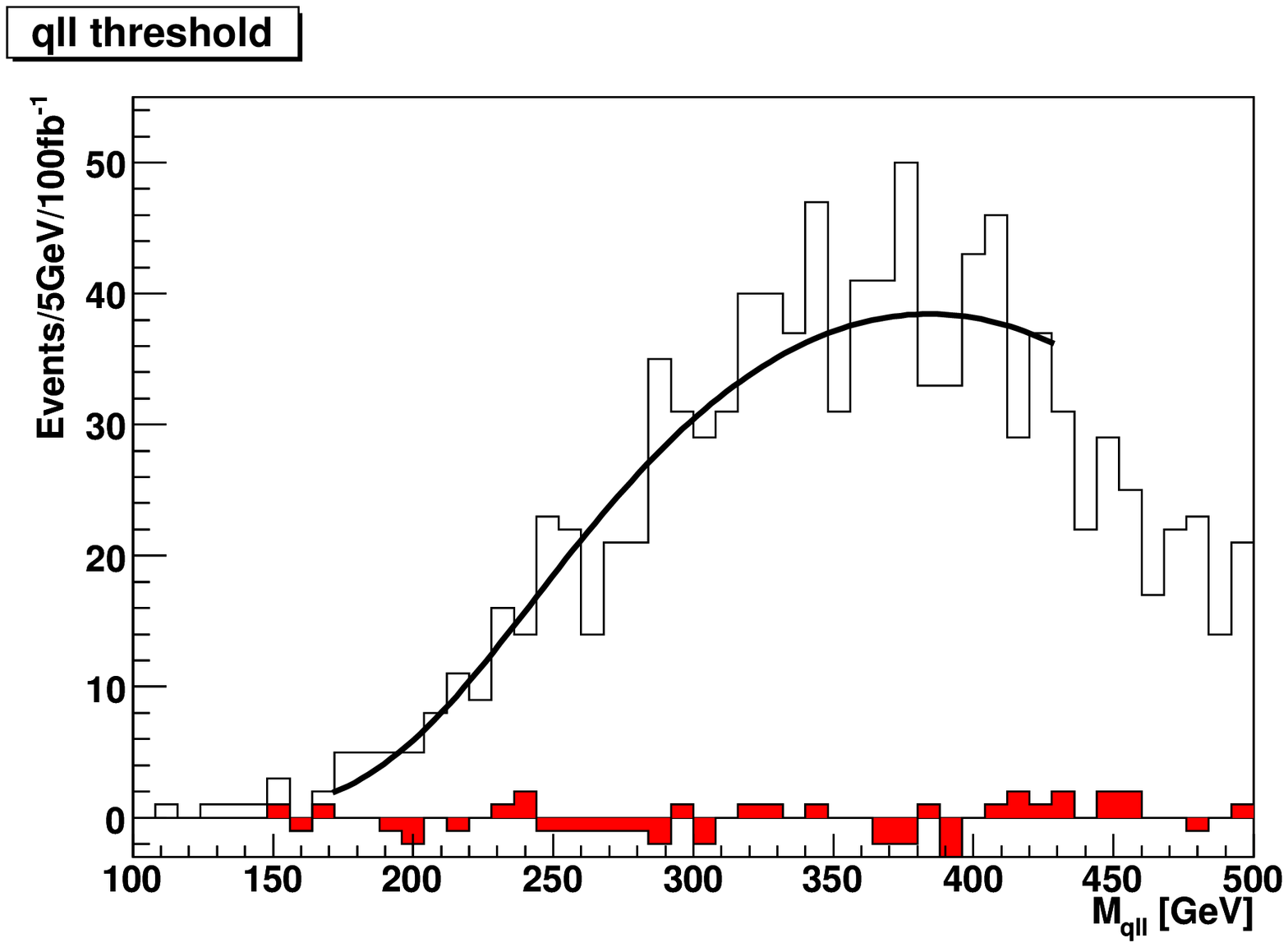}
\caption{Reconstructing the spectrum with a clear di-lepton edge:
  heavier $M_{qll}$ in 4-jet events with $\MET$ and
$M_{ll}>M^{max}_{ll}/\sqrt{2}$. A low-energy threshold is expected. Standard
  Model background is shown in red.}
\label{qllthreshold}
\end{center}
\end{figure}

\begin{figure}
\begin{center}
\includegraphics[width=12cm]{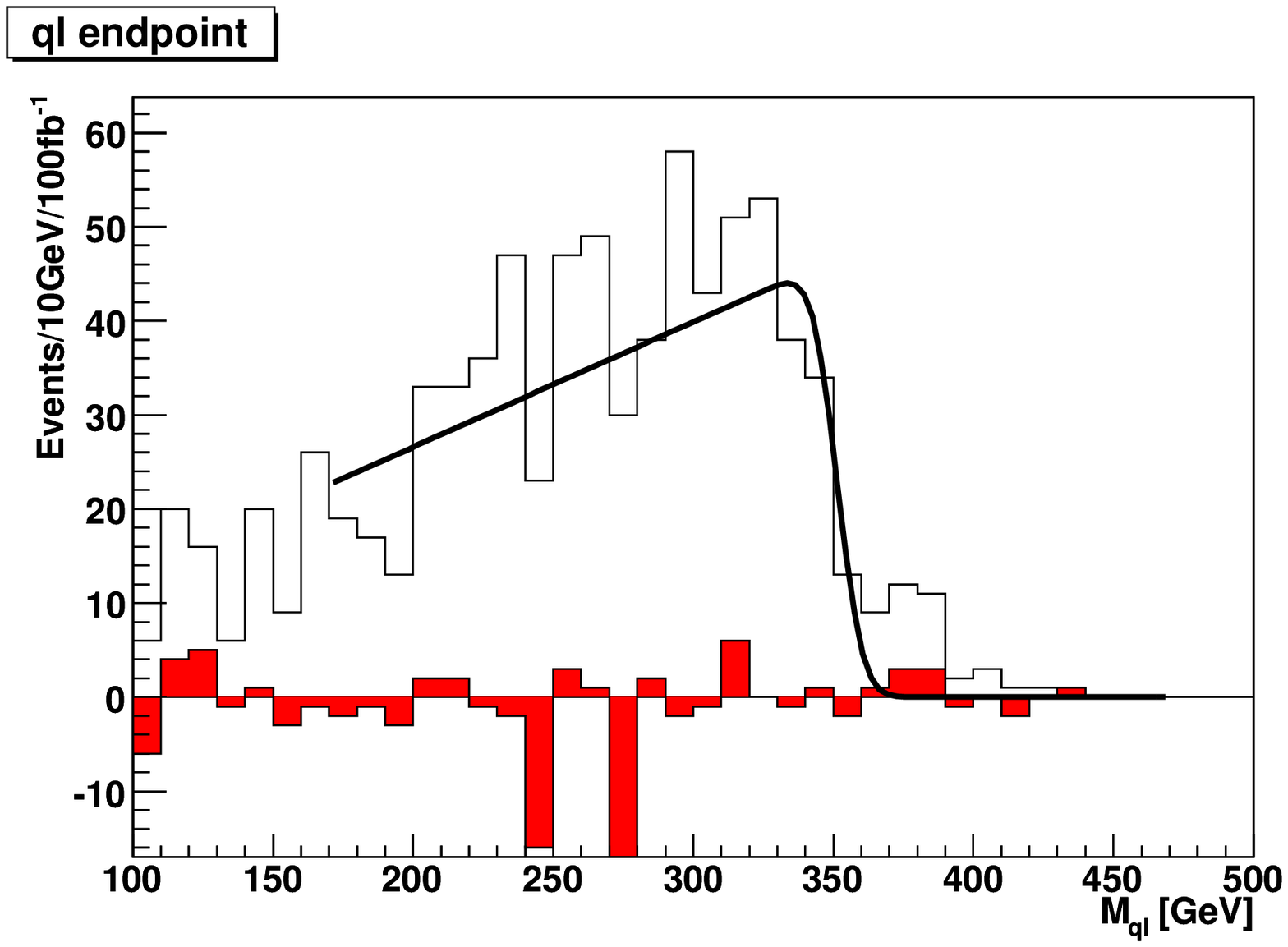}
\caption{Reconstructing the spectrum with a clear di-lepton edge from events
  passing cuts selection B and $M_{qll}<M_{qll}^{max}$: the
  $M_{ql}$ edge. We expect an edge at 371 GeV from the spectrum. Standard Model
  background is shown in red.}
\label{qledge}
\end{center}
\end{figure}

Following Ref.~\cite{ATLASTDR}, we next study the decay channel $\tilde{q}_L
\to q \tilde{\chi}^0_2 \to \tilde{l}^{\pm} l^{\mp} q \to \tilde{\chi}_1^0 l^{\pm}
l^{\mp} q$ to obtain a set of constraints on the squark masses as well as
$\tilde{e}_R$, $\tilde{\chi}^0_1, \tilde{\chi}^0_2.$ Events are selected with
the following properties ({\em selection B}):
\begin{enumerate}
\item
At least 4 hard jets, with $P_{T,1(2)(3)(4)} > 100(50)(50)(50) \hbox{~GeV}$.
\item Two opposite sign electrons or muons
with $P_T>10$ GeV.
\item $M_{eff} \equiv P_{T_1}+P_{T_2}+P_{T_3}+P_{T_4}+\MET \ge 400 \hbox{~GeV}$.
\item
$\MET\ge {\rm max} (300, 0.2M_{eff}).$
\end{enumerate}
The di-lepton 4-momentum is combined
with each of the two hardest jets to obtain two different $qll$ invariant masses. The
lighter of these is plotted in figure \ref{qlledge}.
$M_{llq}^{max}$ can be written in terms of the maximum of several
terms that
contain sparticle masses and have a form similar to
Eq.~\ref{eqll}. In refs.~\cite{hinchliffe,ATLASTDR}, rather simple expressions
  were given for edges such as $M_{llq}^{max}$. These
  were correct for the particular mass spectra examined in those papers, but
  are not true in the general case. We therefore use the general expressions
  given in Ref.~\cite{Allanach:2000kt} and refer the reader there for further
  details.
An empirical fit of the form
\begin{equation}
f(M) = \int^{M_{llq}^{max}}_0 dz (a_1 (M_{llq}^{max}-z) + a_2 (M_{llq}^{max} -
z)^2 )
\exp\left(-{1\over{2\sigma^2}} (M-z)^2\right) + b_1 + b_2 M,
\end{equation}
is used to reconstruct this endpoint with $\sigma=25$ GeV. With a
  variable $\sigma$ in the fit, MINOS did often not converge, indicating
  a possible degenerate $\chi^2$ minimum valley.  With a fixed width, we were
  able to obtain a perfectly good fit, as Fig.~\ref{qllll} shows.
The endpoint obtained through this fit is
$M_{llq}^{max} = 450 \pm 5.5\pm 2.4$ GeV. Where we quote two uncertainties,
the first is a
statistical one from the fitting procedure whereas the second is
our guess at an additional  `fitting' systematic uncertainty from
seeing the effect of changing the bin-size and fit interval.
For events in which $M_{ll}^{max} > M_{ll}>M_{ll}^{max}/\sqrt{2}$, the larger
$qll$ mass is plotted in figure \ref{qllthreshold}.\footnote{The flavour
subtraction of the Standard Model background
in figures \ref{qllll} and \ref{qledge} and can be seen to slightly over subtract. This can
be understood as an artifact of the triggers
used, which have a small flavour asymmetry.}
This gives a threshold which has a theoretical locus in terms of
masses of all four sparticles involved in the chain~\cite{Allanach:2000kt}.
The $qll$ threshold is obtained by fitting with the empirical form
\begin{equation}
[A(M-M_{llq}^{low}) + B (M-M_{llq}^{high})] \theta (M-M_{llq}^{high}),
\end{equation} and a Gaussian smearing.
We obtain $M_{qll}^{min} = 188\pm 7.5\pm 6.6$ GeV from the fit, which
Fig.~\ref{qllthreshold} shows, reproducing the shape of binned simulated data
up to statistical variations.

Events are then further selected
such that one $qll$ mass is less than and the other greater
than $M_{qll}^{max} \sim 500$ GeV. This identifies the jet involved
in the decay chain $\tilde{q}_L \to
q \tilde{\chi}^0_2 \to \tilde{l}^{\pm} l^{\mp} q \to \tilde{\chi}_1^0 l^+
l^- q$.
By combining each of the leptons with this jet, we can plot the $ql$ mass
(figure \ref{qledge}) which constrains a function of the sparticle
masses~\cite{Allanach:2000kt}.
This endpoint is located in a similar fashion to the di-lepton endpoint, using
a Gaussian smeared triangular fit. We obtained $M_{lq}^{max} = 353\pm
1.7\pm 3.6$ GeV.

We summarise the values found through the above fitting procedure in
Table~\ref{tab:fit}. They differ from the expected
values
by `experimental' systematic errors. The sources for these systematic errors
can be associated with hadronic
calorimeter calibration, jet energy leakage or the cone or cluster algorithms that reconstruct the jets.
For our purposes we assume that these systematic errors can be
removed as part of the experimental analysis. Following \cite{hinchliffe},
for our estimation of sparticle masses we shift the measured endpoints to the
theoretical values,
thus eliminating this experimental systematic bias, but keep the errors
arising from the fitting procedure. We add the statistical and `fitting'
systematic errors in quadrature in order to quote a total uncertainty on the
central theoretical value.
\TABULAR[t]{|c|c|c|}{\hline
 & Fitted value/GeV & Shifted value/GeV \\ \hline
$M_{ll}^{max}$ & 69.4$\pm$ 0.15  & 69.4$\pm$0.15  \\
$M_{llq}^{max}$ & 450$\pm$5.5$\pm$2.4 & 467.6$\pm$6.0  \\
$M_{lq}^{max}$ & 353$\pm$1.7$\pm$3.6 & 370.8$\pm$4.0\\
$M_{llq}^{min}$ & 188$\pm$7.5$\pm$6.6 & 202.8$\pm$10.0\\ \hline
}{Fitted edges and their uncertainties. \label{tab:fit}}

\begin{figure}
\threegraphs{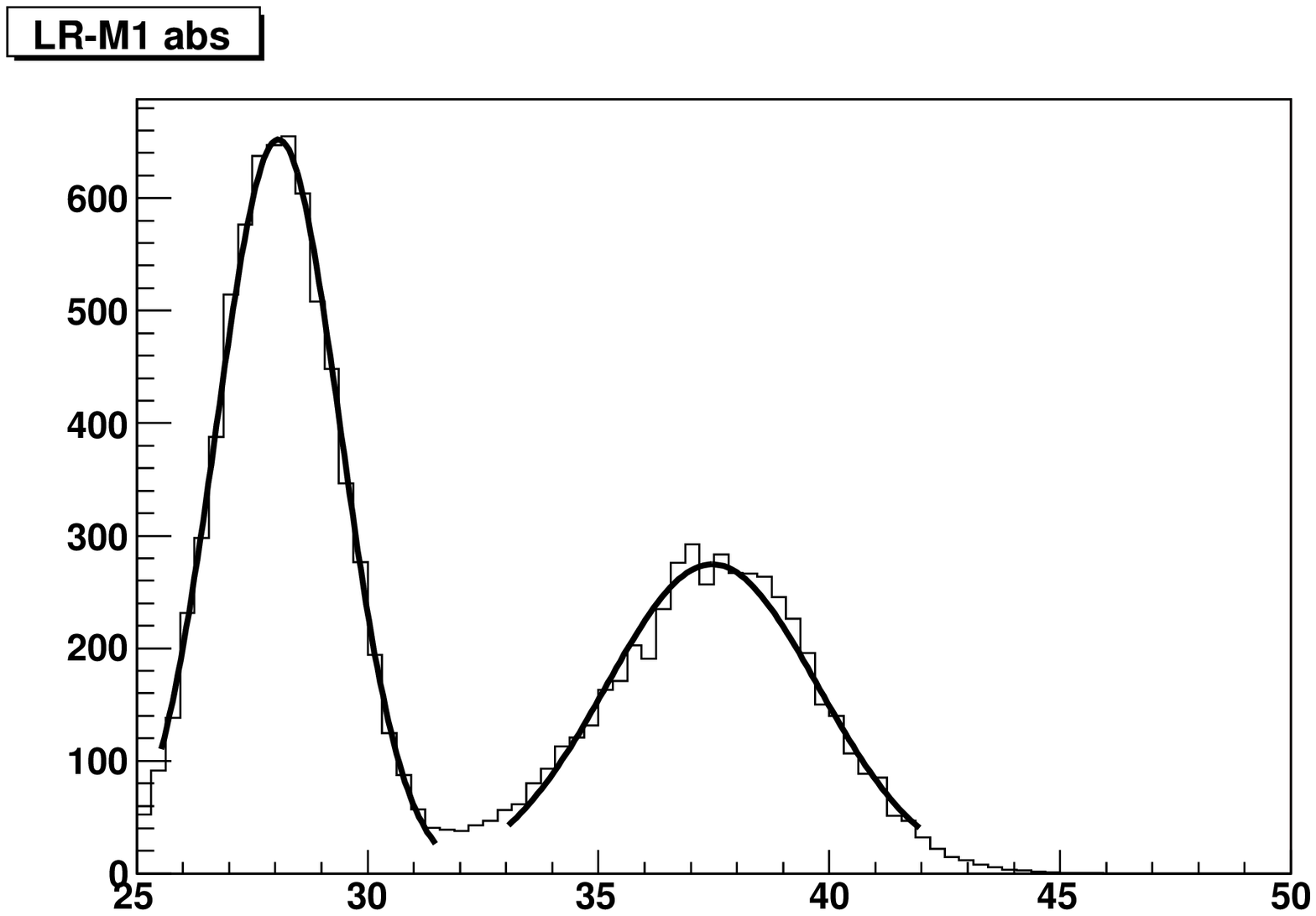}{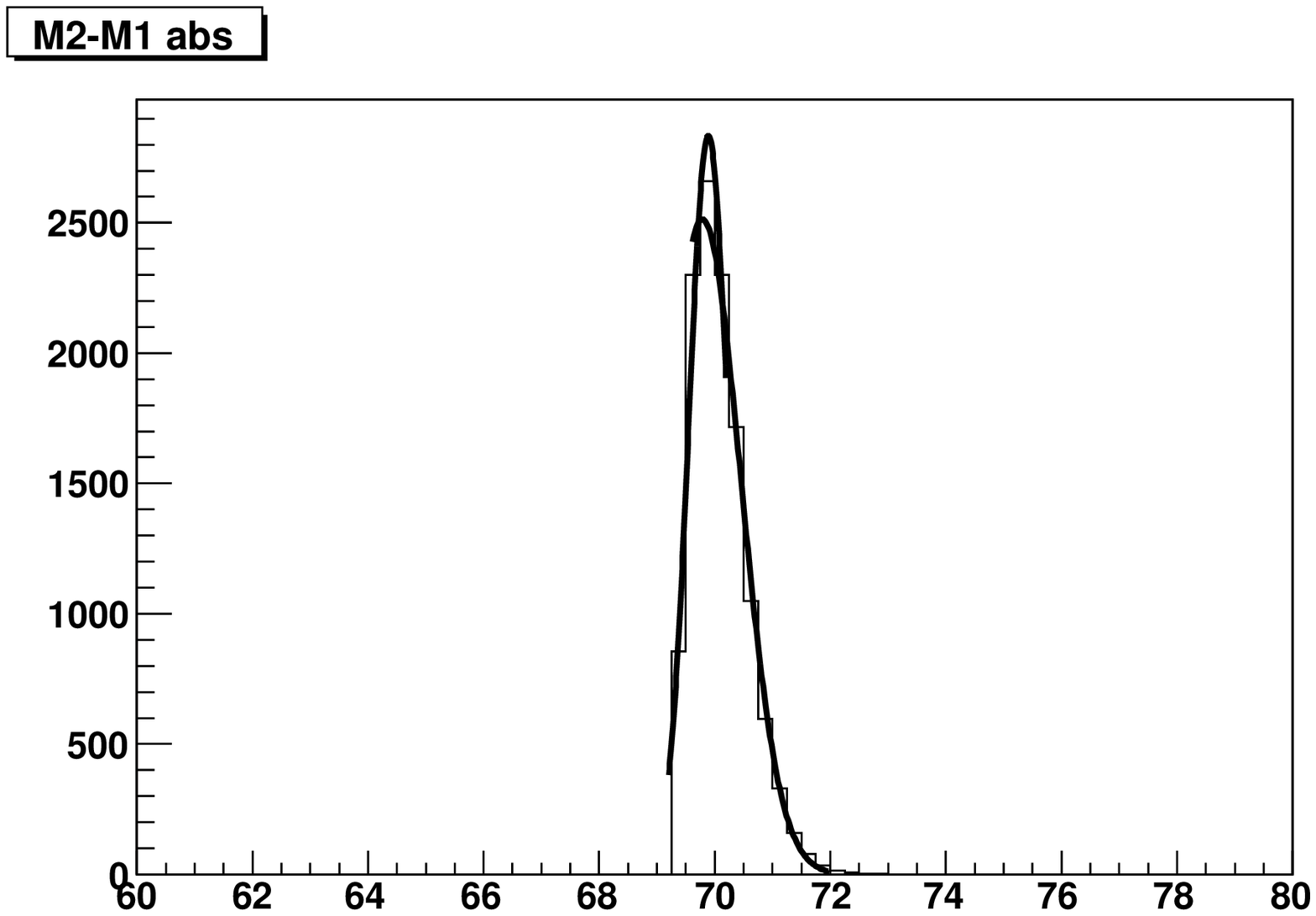}{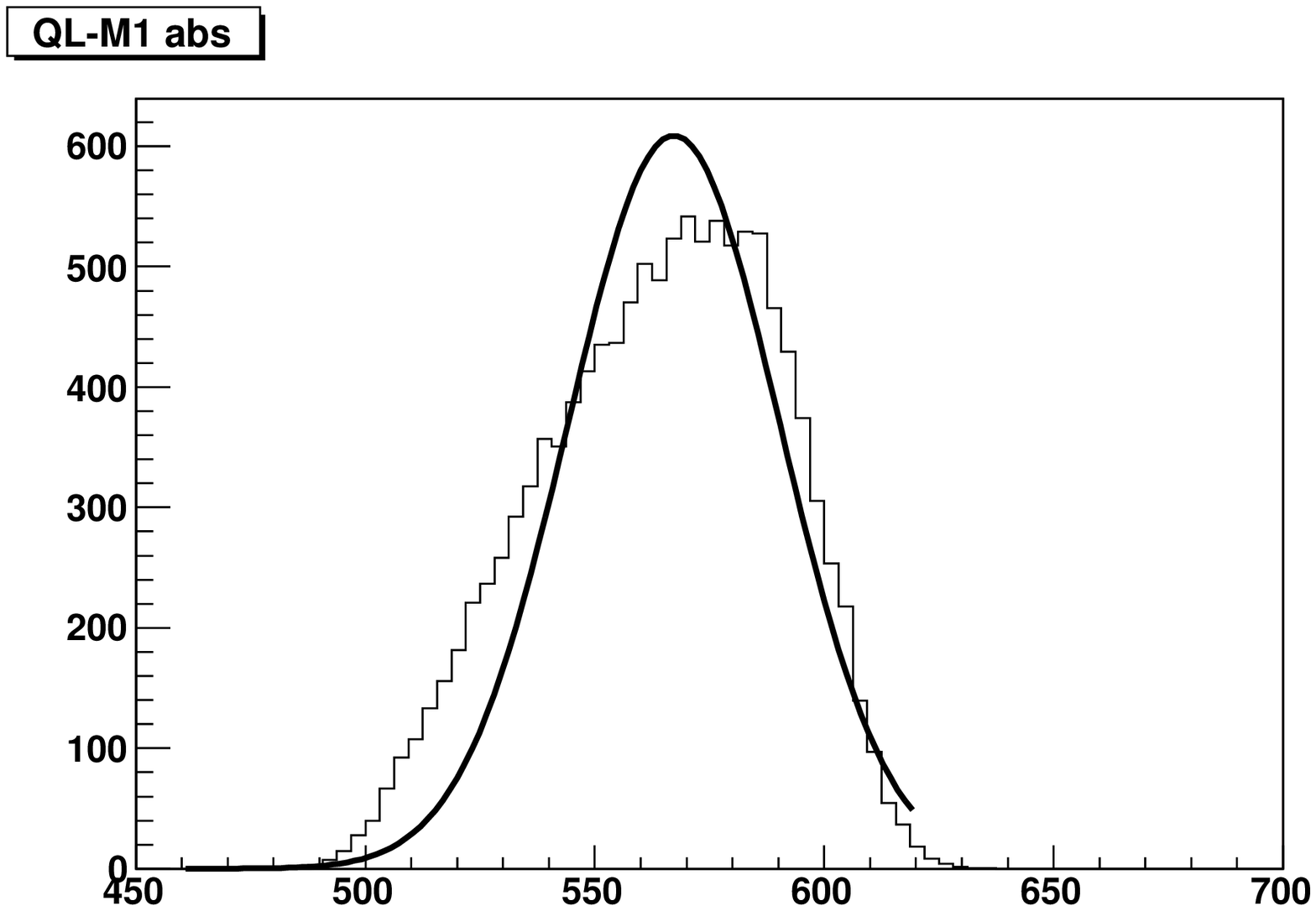}
\caption{Mass difference samplings for the reconstructed spectrum with $100
    {\rm
    fb}^{-1}$ of data:
(a) $m_{\tilde{l}_R} - m_{\tilde{\chi}_1^0}$, (b) $m_{\tilde{\chi}_2^0} -
  m_{\tilde{\chi}_1^0}$, (c) $m_{\tilde{q}_L} - m_{\tilde{\chi}_1^0}$.\label{massdiffs100fb}}
\end{figure}

We now use these shifted endpoints to reconstruct the $\tilde{\chi}_1^0,
\tilde{q}_L, \tilde{e}_R$ and $\tilde{\chi}_2^0$
masses.
To do so, we take $m_{\tilde{q}_L}, m_{\tilde{e}_R}$ and $m_{\tilde{\chi}^0_2}$
to be randomly generated with a uniform distribution within 50\% of
their central values and compute $m_{\tilde{\chi}^0_1}$ using the
$M_{ll}^{max}$ di-lepton endpoint which has a very small statistical
error. This is equivalent to approximating the very narrow Gaussian likelihood
distribution of $M_{ll}^{max}$ with a $\delta$ function.
We then compute the $\chi^2$ for the remaining observables ($M_{llq}^{max},
M_{llq}^{min},
M_{lq}^{max}$) and assign a weight $\propto e^{-\chi^2/2}$ to
this set of randomly generated masses. Doing this many times
provides a sampling of a probability distribution for the sparticle masses.
The marginalisations to three independent mass differences
is shown in figure \ref{massdiffs100fb}.
\TABULAR[t]{|c|cc|}{\hline
 & Theoretical/GeV & Estimated/GeV \\ \hline
$m_{\tilde{l}_R} - m_{\tilde{\chi}_1^0}$ & 37.5& 28.1$\pm$1.4, 37.5$\pm$2.3  \\
$m_{\tilde{\chi}_2^0} -  m_{\tilde{\chi}_1^0}$ &69.7 & 69.7$^{+0.7}_{-0.3}$ \\
$m_{\tilde{q}_L} - m_{\tilde{\chi}_1^0}$ & 567& 564$\pm$26\\ \hline
}{Sparticle mass differences. \label{tab:dm}}
From each probability distribution, we estimate the mass differences.  The
    histograms show the binned estimated probability density function of each
    mass difference, whereas the continuous lines show the best-fit Gaussian
    shape. In Fig.~\ref{massdiffs100fb}a, the plot has been separated into two
 regions, each of which is fitted with a separate Gaussian. $m_{{\tilde
 \chi}_2^0}-m_{{\tilde \chi}_1^0}$ has been fitted with two half-Gaussians of
 different widths, ``glued'' at the maximum.
 $m_{{\tilde q}_L}-m_{{\tilde \chi}_1^0}$ is not very well fitted with a
 Gaussian,  as  Fig.~\ref{massdiffs100fb}c shows. We characterise the
 distribution instead  by its mean and standard deviation.
The characteristic double-bump structure of Fig.~\ref{massdiffs100fb}a
 displays the existence of two solutions in mass space for the edge variables
 listed in Table~\ref{tab:fit} and results in two different possible estimated
 values for the mass difference, one at each local maximum.
This ambiguity was also observed for the case of an LHC SPS1a spectrum
 reconstruction in \cite{miller}. We display the estimated and theoretical
 values for the three mass differences in Table~\ref{tab:dm}.
In our case, the 'true' peak correspond to
  the ql-edge given by invariant mass of the quark and the far lepton in the
  decay chain, whereas the 'wrong' peak correspond to a solution for the
  ql-edge given by the quark and the near lepton.
We find the `wrong' solution by numerical scans to be
$m_{\tilde{q}_L} = 792.6$ GeV, $m_{\tilde{\chi}_2^0} = 303.6$ GeV,
$m_{\tilde{l}_R} = 261.8$ GeV and $m_{\tilde{\chi}_1^0} = 233.7$ GeV with a
 resulting $\chi^2$
 of 0 (the solution of the correct peak also has a zero value of $\chi^2$, by
 construction). The different heights of the bumps in
 Fig.~\ref{massdiffs100fb} must therefore be a consequence of volume effects
 in the  marginalisation procedure,
 since the two best-fit solutions are equally likely.
One requires additional data in order to discriminate between the two
 solutions experimentally.

Despite the existence of two possible good-fit regions of mass difference
space, we have enough information to discriminate against mSUGRA
models.
As mentioned before, the gaugino mass ratios
in mSUGRA are $M_1:M_2:M_3 \approx 1:2:6.$
We also know that in mSUGRA, $m_{\tilde{q}_L}$ is strongly correlated with $m_{\tilde{g}}$ and
generically we have $m_{\tilde{q}_L} \lesssim m_{\tilde{g}}$ (this is
guaranteed
within mSUGRA by the presence of light sleptons).
Therefore, in mSUGRA models we have
$m_{\tilde{g}} \approx 6m_{\tilde{\chi}_1^0}$ and
\bea
(m_{\tilde{g}} - m_{\tilde{\chi}_1^0})/
(m_{\tilde{\chi}_2^0} -  m_{\tilde{\chi}_1^0}) & = & 5, \nonumber \\
(m_{\tilde{g}} - m_{\tilde{\chi}_1^0})/
(m_{\tilde{\chi}_2^0} -  m_{\tilde{\chi}_1^0}) & \gtrsim & (m_{\tilde{q}_L} - m_{\tilde{\chi}_1^0})/
(m_{\tilde{\chi}_2^0} -  m_{\tilde{\chi}_1^0}). \nonumber
\eea
However, for the large volume models we obtain
\be
(m_{\tilde{q_L}} - m_{\tilde{\chi}_1^0})/
(m_{\tilde{\chi}_2^0} -  m_{\tilde{\chi}_1^0}) = 8.09\pm 0.38.
\ee
from Table~\ref{tab:dm}.
Thus the measurements made are not compatible with
the mSUGRA scenario. Further measurements, e.g. of the gluino mass
and the right handed squark masses, would provide further evidence
for discrimination against mSUGRA. Using the expected $m_{\tilde{q_L}}:m_{\tilde{g}}$ ratios,
we could also investigate the $M_2:M_3 \approx 1:3$ prediction of the large volume models.
For this it would be useful to directly measure the gluino mass: to this end
the decay channel $\tilde{g}\to q \tilde{q}_R \to q q \tilde{\chi}_1^0$ may be exploited as well as the
$M_{T_2}$ variable \cite{MT2}.

\section{Conclusions}
\label{conclusions}

We have performed a detailed study of the expected superparticle spectrum and
collider phenomenology for large-volume string models.
Our main conclusions are:

\begin{enumerate}

\item
The large volume models give rise to a distinctive spectrum of gaugino masses, characterised by
\be
M_1 : M_2 : M_3 = (1.5 \to 2) : 2 : 6
\ee
This can be distinguished from the ratios that appear in e.g. mirage mediation or mSUGRA.

The collider phenomenology depends heavily on the mass difference between
$M_1$ and $M_2$ and the slepton mass spectrum. If this is large,
leading to many $\tilde{\chi}_2^0 \to \tilde{\chi}_1^0 l^{\pm} l^{\mp}$ events, kinematical reconstruction of the spectrum is
much easier. This was discussed in section \ref{observables}.

\item
The overall spectrum tends to be more bunched than that of a
corresponding mSUGRA model. This can be understood by the approximate unification, prior to the
inclusion of the effects of magnetic fluxes on the brane world-volume, of
scalar and gaugino masses at the intermediate (fundamental) scale. There is
then less energy for the physical masses to evolve from their theoretical
boundary condition and the overall spectrum
falls within a narrower mass range.

This effect also occurs in models of mirage mediation, where gaugino masses
are (accidentally) unified at the intermediate scale.

\item
More concretely we find: the LSP is mostly bino. The second neutralino is
mostly wino and is almost degenerate with charginos.
Sleptons are almost degenerate, with
stau the lightest. The gluino is the heaviest sparticle. The ratio of
the gaugino-squark masses is larger than that predicted by mSUGRA.

\item

We have quantified the uncertainty that appears in the weak scale spectra due
to uncertainties in the
high-energy soft terms. The incorporation of such uncertainties is essential
in trying to make predictions for LHC signatures
based on high-scale string constructions.

\item
We have used event generators and detector simulators to study
possible signatures of our models. We analysed the use and limitations of
certain
`counting' observables to contrast our models with other classes of
models, especially a line through mSUGRA parameter space (the SPS1a slope)
that has been well studied
in the literature. We found that some counting observables are more
useful than others but in general they do not provide enough
information  to fully distinguish  the models, at least in simple
2-dimensional projections. It may be true that a fit of the models to the full
parameter space of counting observables is required.

\item
We studied in detail a sample model that is quite rich in $\tilde{\chi}_2^0 \to
\tilde{\chi}_1^0 l^{\pm} l^{\mp}$ decays.
Accurate reconstruction of some  properties of the low-energy
sparticle spectrum is possible with 100 fb$^{-1}$ of integrated luminosity.
Our sample study shows that
the large volume model can be differentiated from standard mSUGRA.

\end{enumerate}

In this work we have made progress in
the process of starting with a well defined class of microscopic
models and bringing them to the point where they may be confronted
with potential experimental measurements at the LHC.
This is a positive step in the direction of testing
classes of models derived from string theory.
This is clearly a less ambitious task than testing string theory
in its entirety, but one that may prove more fruitful. Moduli
stabilisation with supersymmetry breaking has allowed us to
find explicit expressions for soft breaking terms that have a well-defined
microscopic origin, which is not the case for the well studied
standard benchmark points. Issues, such as flavour universality and extra CP
violation, that render the generic gravity mediation scenario
unrealistic and have to be resolved by hand in most models, can now be
understood in terms of the particular properties of string compactifications. 
 Furthermore we have found
ways to distinguish our models from mSUGRA and other models, using properties that may be
directly measured within a
few years of LHC running.

There are however several open questions we need to emphasise. First
of all we have been working on a scenario allowed by type II
constructions, in which the Standard Model is assumed to live on a set
of D-branes localised at a particular region inside the Calabi-Yau manifold.
The problems of moduli stabilisation and supersymmetry breaking are
thus decoupled from the details of the Standard Model
construction. This approach
has the positive feature that moduli are stabilised
in a large class of models. Other issues,
such as the number of families, proton stability and gauge
unification are more model dependent. In this sense our results are
very robust. On the other hand they lack concreteness in the sense that
we do not have an explicit D-brane configuration with the MSSM
spectrum, known Yukawa couplings, etc. Finding a fully realistic model
in this approach is therefore an open question.

Nevertheless we have
identified the main sources of uncertainty in our analysis that
parametrise our ignorance of such a realistic model. First, we
included the effects of magnetic fluxes, usually needed to construct
chiral  models on D7 branes. Even though
the dependence of K\"ahler potentials and gauge couplings on the
magnetic fluxes is not known, we
were able to parametrise our ignorance in terms of the random
parameters $\epsilon$. A second source of uncertainty is the spectrum
itself. We know that typical quasi-realistic  D-brane models (see
\cite{marchesano} for a recent review) usually have extra particles
beyond the MSSM and that the hypercharge does not have a canonical
normalisation. We took into account these effects by varying
the hypercharge normalisation and finding observables, such as the
ratios of gaugino masses, which do not change if there are
extra fields beyond the MSSM. Finally, we have assumed the simplest
configuration of D7 branes hosting the Standard Model in the sense
that they are all assumed to wrap the same 4-cycle. Different
configurations would slightly change the expressions for the soft
breaking terms. The expressions would then depend on an extra parameter $\lambda$ that takes
different values depending upon which cycles the branes wrap and the manner of their intersection.
For the case studied here, $\lambda = 1/3$ \cite{hepth0609180,hepth0610129}.
It would be interesting to explore
 the implications of other configurations leading to different
 values of $\lambda$.

The model independence of our analysis makes it easy to adapt once
explicit realistic models are constructed  that may differ
from the MSSM. Furthermore,
potential experimental measurements at the LHC may provide guidance on
what the structure of these realistic models should be.  Even if at
the end
it turns out that our
models will not pass experimental scrutiny from the LHC, the detailed analysis
made, all the way from string theory to LHC observables, should be
a useful guide for future proposals. It is
encouraging to have this rich interplay between theory and experiment
waiting for the arrival of LHC results.

\section*{Acknowledgements}
This work has been partially supported by STFC.\footnote{Formerly PPARC
  (R.I.P.)}
We thank Shehu Abdussalam, Marcus Berg, Kiwoon Choi,
Daniel Cremades
and the members of the Cambridge SUSY working group for
useful conversations. We particularly thank our colleagues that wrote
and made available
the many codes we have used in this work.
The computational work has been performed using the Cambridge eScience CAMGRID
computing facility, with the invaluable help of Mark Calleja and Andy Parker.
JC and KS are funded by Trinity College, Cambridge.
BCA is funded by STFC. FQ
is partially funded by STFC and a Royal Society Wolfson award.  CHK is
supported by a Dorothy Hodgkin Postgraduate Award.

\appendix
\section{PGS Level 2 Triggers}
Here we list the PGS Level 2 Triggers used in our event analysis for ease of
reference, which can be obtained from pgs\_olympics.f in the PGS package
\cite{PGS}. If any of the following apply to an event, the trigger is passed and the event recorded.
The PGS definition of isolation is somewhat involved, and we refer the
interested reader to the PGS manual for further details. Our
isolation criteria for muons are based on those used in the {\tt Chameleon} \cite{chameleon}
package: the ratio of the energy in a 3x3 grid surrounding
the muon and the $p_T$ of the muon is required to be $<$ 0.1125, and
the total $p_T$ of a $\Delta R = 0.4$ cone region surrounding
(but excluding) the muon is required to be $< 5$ GeV.

\begin{enumerate}
\item
Inclusive isolated lepton $l\equiv e,\mu$
$p_T(l)> 180$ GeV;
\item
For a lepton $p_T(l)>130$ GeV plus a jet $j$
$p_T(j)>200$ GeV;
\item
Isolated same-flavour di-leptons $p_T (l_{1,2})>60$ GeV;
\item
Di-leptons $p_T (l_{1,2})>45$ GeV plus jet $p_T(j)>150$ GeV;
\item
Isolated opposite-flavour di-leptons $p_T (l_{1,2}) >30$ GeV;
\item
Isolated lepton $p_T(l)>45$ GeV plus isolated tau $p_T(\tau)>60$ GeV;
\item
Isolated di-tau $p_T (\tau_{1,2}) >60$ GeV;
\item
Inclusive isolated photon $p_T(\gamma)>80$ GeV;
\item
Isolated di-photon $p_T (\gamma_{1,2}) > 40$ GeV;
\item
Inclusive $\MET> 200$ GeV;
\item
Inclusive single-jet $p_T>1000$ GeV;
\item
Jet $p_T(j)>300$ GeV plus $\MET>125$ GeV;
\item
Acoplanar jet $p_T(j)>150$ GeV and $\MET>80$ GeV, $1 < \Dphi_{j \,  \, \,  {\MET}}
< 2$;
\item
Acoplanar dijets $p_T (j_{1,2}) >400$ GeV, $\Dphi_{jj} < 2$.
\end{enumerate}

\end{document}